Grant Agreement No.: 671705
Research and Innovation action
Call Identifier: H2020-ICT-2014-2

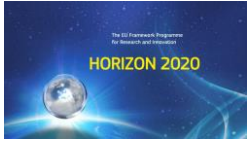
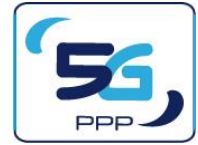

# Speed5G

**quality of Service Provision and capacity Expansion through Extended-DSA for 5G**

# D3.2: SPEED-5G enhanced functional and system architecture, scenarios and performance evaluation metrics

Version: v1.2

| Deliverable type | Report |
|---|---|
| Dissemination level | PU (Public) |
| Due date | 30/06/2016 (revision: 30/11/2016) |
| Submission date | 30/06/2016 (revision: 30/11/2016) |
| Lead editor | Shahid Mumtaz (IT) |
| Authors | Shahid Mumtaz, Kazi Saidul Huq, Jonathan Rodriguez, Paulo Marques, Ayman Radwan (IT), Keith Briggs & Michael Fitch (BT); Andreas Georgakopoulos, Ioannis-Prodromos Belikaidis, Panagiotis Vlacheas, Dimitrios Kelaidonis, Evangelos Kosmatos, Serafim Kotrotsos, Stavroula Vassaki, Yiouli Kritikou, Panagiotis Demestichas, Kostas Tsagkaris, Evangelia Tzifa, Aikaterini Demesticha, Vera Stavroulaki, Athina Ropodi, Evangelos Argoudelis, Marinos Galiatsatos, Aristotelis Margaris, George Paitaris, Dimitrios Kardaris, Ioannis Kaffes (WINGS); Haeyoung Lee, Klaus Moessner (UNIS); Valerio Frascolla, Bismark Okyere (Intel); Salva Díaz, Óscar Carrasco, Federico Miatton (Sistel); Antonio de Domenico, Benoit Miscopein (CEA); Thanasis Oikonomou, Dimitrios Kritharidis (ICOM); Harald Weigold (R&S) |
| Work package, Task | WP 3, T3.2, T3.3, T3.4 |
| Keywords | 5G, Scenario, Use case, Architecture, Virtualisation, KPIs |

*Abstract:* This deliverable contains a detailed description of the use cases considered in SPEED-5G, which will be used as a basis for demonstration in project. These use cases are Dynamic Channel selection, Load balancing, carrier aggregation. This deliverable also explains the SPEED-5G architecture design principles, which is based on software-defined networking and network function virtualisation. The degree of virtualisation is further illustrated by a number of novel contributions from involved partners. In the end, KPIs for each use case are presented, along with the description of how these KPIs can support 5G-PPP KPIs.

**Document revision history**

| Version | Date | Description of change | List of contributor(s) |
|---------|------|----------------------|------------------------|
| 0.1 | 2015-12-07 | Initialisation of ToC | IT |
| 0.2 | 2016-01-10 | Contribution on virtualization | SISTEL, IT, WINGS,BT, Intel, UNIS |
| 0.2 | 2016-02-20 | Revision of virtualisation | SISTEL, IT, WINGS,BT, Intel, UNIS |
| 0.3 | 2016-02-28 | Initial contribution on KPI | UNIS, IT, WINGS,BT, R&S, ICOM |
| 0.3 | 2016-03-25 | Discussion on KPI | UNIS, IT, WINGS,BT, R&S, ICOM |
| 0.4 | 2016-04-20 | Initial proposal on use cases | SISTEL, CEA, IT, WINGS,BT |
| 0.5 | 2016-05-25 | Finalised the use cases | SISTEL, CEA, IT, WINGS,BT |
| 0.6 | 2016-06-10 | Integrating all contribution | IT |
| 0.7 | 2016-06-25 | Editing and Revision of D3.2 | IT |
| 1.0 | 2016-06-27 | review | BT |
| 1.1 | 2016-11-10 | New sections 3.1.2 and 3.1.3 | SISTEL |
| 1.2 | 2016-11-30 | New section 3 and updates in other sections | UNIS, IT, WINGS,BT, R&S, ICOM |

---

[1] http://creativecommons.org/licenses/by-nc-nd/3.0/deed.en_US

# Executive Summary

SPEED-5G's main objective is to investigate and develop technologies that address the well-known challenges of predicted growth in mobile connections and traffic volume by successfully addressing the lack of dynamic control across wireless network resources, which is leading to unbalanced spectrum loads and a perceived capacity bottleneck. As a result, SPEED-5G focuses on enhanced dynamic spectrum access (eDSA) with three degrees of freedom: densification, rationalised traffic allocation over heterogeneous wireless technologies, and better load balancing across available spectrum.

To tackle the eDSA problem, SPEED-5G considered four scenarios, having been discussed in D3.1. These scenarios covered the SPEED-5G concepts for urban communications, IoT, mission-critical and vehicular services and as a result are representative for validating RRM/MAC solutions and eDSA techniques that SPEED-5G will propose. This deliverable further enhances the eDSA scenarios into the following three realistic and demonstrable use cases.

- Dynamic channel selection

- Load balancing

- Carrier aggregation

Details of each of the use cases, refinement of function processes in terms of RRM/MAC will be explained in WP4 (D4.1) and WP5 (D5.1). This deliverable also presents the SPEED-5G architecture design principles which are based on flexibility, simplicity, on-demand resource allocation, auto-scaling and enhanced performance through extensive utilisation of software-defined networking and network function virtualisation. The degree of virtualisation is further elaborated in terms of virtual small cell deployment, virtualisation of fronthaul/backhaul, virtualisation effect on RRM/MAC, and KPIs to measure virtualisation and its standardisation. Finally, this deliverable explains various KPIs for different SPEED-5G use cases and scenarios, along with the relation between SPEED-5G KPIs and 5G-PPP KPIs.

This document will be used in future SPEED-5G WPs to implement the target requirements. It also serves as a reference for the project evaluation where achieved performance will be benchmarked against stated targets.

# Table of Contents

# List of Figures

# List of Tables

# Abbreviations

| ACK | Acknowledge |
|-----|-------------|
| ACS | Auto-Configuration Server |
| AL | Adaptation Layer |
| AM | Acknowledge Mode |
| AP | Access Point |
| ARPU | Average Revenue Per User |
| BBU | Baseband Unit |
| BER | Bit Error Rate |
| BLER | Block Error Rate |
| BW | Bandwidth |
| CAPEX | Capital Expenditure |
| CCA | Clear Channel Assessment |
| CCDF | Complementary Cumulative Distribution Function |
| CDF | Cumulative Distribution Function |
| CPRI | Common Public Radio Interface |
| CIR | Carrier-to-Interference Ratio |
| CN | Core Network |
| CNR | Carrier-to-Noise Ratio |
| CoMP | Coordinated Multi-Point |
| CPRI | Common Public Radio Interface |
| CQI | Channel Quality Indicator |
| cRRM | Centralised Radio Resource Management |
| CSI | Channel State Information |
| D2D | Device-to-Device Communication |
| DASH | Dynamic Adaptive Streaming over HTTP |
| DLC | Data Link Control |
| DTV | Digital TeleVision |
| DUR | Desired-to-Undesired power Ratio |
| EC | European Commission |
| eDSA | Enhanced Dynamic Spectrum Access |
| eMBB | Enhanced Mobile Broadband |
| FDD | Frequency Division Duplex |
| FTP | File Transfer Protocol |
| FTTC | Fibre to the Cabinet |
| FTTP | Fibre to the Premises |
| GRGR | Global Regulator Repository |
| GW | Gateway |
| HARQ | Hybrid Automatic Repeat Request |
| **HeMS** | HeNB Management System |
| **HTML** | Hypertext Markup Language |
| **HTTP** | Hypertext Transfer Protocol |
| **HW** | Hardware |
| **ICIC** | Inter Cell Interference Cancellation |
| **IEEE** | The Institute of Electrical and Electronics Engineers |
| **InP** | Infrastructure Provider |
| **IoT** | Internet-of-Things |
| **IPTV** | Internet Protocol Television |
| **KPI** | Key Performance Indicator |
| **LAA** | Licence-assisted access |
| **LIPA** | Local IP Access |
| **LNA** | Low-Noise Amplifier |
| **LSPC** | Local Spectrum Control |
| **LTE** | Long Term Evolution |
| **LTE-A** | LTE-Advanced |
| **LTE-U** | LT- Unlicensed |
| **MAC** | Medium Access Control |
| **mIoT** | Massive IoT communications |
| **MME** | Mobility Management Entity |
| **MMT** | MPEG Media Transport |
| **MOCON** | Multi-Operator Core Network |
| **MT** | Mobile Terminal |
| **MU-MIMO** | Multi-User Multiple-Input Multiple-Output |
| **MVNO** | Mobile Virtual Network Operator |
| **NACK** | Negative Acknowledgment |
| **NC** | Network Cognition |
| **NFV** | Network Functions Virtualisation |
| **NFV ISG** | NFV Industry Specification Group |
| **NV** | Network Virtualisation |
| **OAM** | Operations and Management |
| **Ofcom** | UK Office of Communications |
| **OPEX** | Operational Expenditure |
| **PCI** | Peripheral Component Interconnect |
| **PDCP** | Packet Data Convergence Protocol |
| **PDF** | Probability Density Function |
| **PHY** | Physical Layer |
| **PMF** | Probability Mass Function |
| **PMSE** | Program Making and Special Events |
| **QAM** | Quadrature-Amplitude Modulation |

| QoE | Quality of Experience |
|------|------|
| QoS | Quality of Service |
| RAN | Radio Access Network |
| RAT | Radio Access Technology |
| RB | Resource Block |
| RF | Radio Frequency |
| REM | Radio Environmental Map |
| RLC | Radio Link Control |
| RM | Resource Manager |
| RNC | Radio Network Controller |
| RRC | Radio Resource Control |
| RRM | Radio Resource Management |
| SAN | Spectrum Analyser |
| SC | Small Cell |
| SDN | Software-Defined Networking |
| SDR | Software-Defined Radio |
| SINR | Signal-to-Interference plus Noise Ratio |
| SLA | Service Level Agreement |
| SM | Spectrum Manager |
| SNR | Signal-to-Noise Ratio |
| SON | Self-Organizing Network |
| SP | Service Provider |
| SPI | Spectrum Efficiency Index |
| SPRR | Spectrum Provider Repository |
| SS | Spectrum Sensing |
| SSE | Spectrum Selector |
| SW | Software |
| TCP | Transmission Control Protocol |
| TDD | Time Division Duplexing |
| TDMA | Time Division Multiple Access |
| TTI | Transmission Time Interval |
| TVWS | TV White Space |
| UC | Use Case |
| UE | User Equipment |
| URC | Ultra-Reliable Communications |
| UWB | Ultra-Wideband |
| VM | Virtual Machine |
| VoIP | Voice over IP |
| WAP | Wireless Application Protocol |
| WLAN | Wireless Local Area Network |

| **WRA** | Weighted Randomised Algorithm |
|---------|-------------------------------|

# 1    Introduction

Requirements for 5G systems refer to a 1000 times higher capacity, 1ms maximal latency, seamless connectivity across different access technologies and the highest possible Quality of Experience (QoE) for users [1]. To meet these targets, significant operational and infrastructure investments will be required. Unfortunately, revenues are not growing at the same rate as costs, as average revenue per user (ARPU) is expected to remain at best constant in mature markets. Moreover, the current lack of dynamic control across wireless network resources is leading to unbalanced spectrum loads and a perceived capacity bottleneck.

Tackling inefficiencies is therefore one key aspect of the SPEED-5G approach. A major challenge for future networks is to map various types of traffic and Quality of Service (QoS) requirements across the most appropriate radio technologies and spectrum bands, taking into consideration a larger variety of licensing schemes compared with current deployments.

Therefore, SPEED-5G's main objective is to achieve a significantly better exploitation of heterogeneous wireless technologies, providing higher capacity together with the ultra-densification of cellular technology and effectively supporting the new 5G QoE requirements. As a result, SPEED-5G focuses on enhanced dynamic spectrum access (eDSA) with three degrees of freedom: densification, rationalised traffic allocation over heterogeneous wireless technologies and better load balancing across available spectrum.

SPEED-5G scenarios for urban communications, IoT, mission-critical, and vehicular services are described in D3.1. This deliverable proposes three related uses cases for validating proposed RRM/MAC solutions in SPEED-5G and elaborates on the impact (and degrees of) of virtualization on the proposed framework.  .

The remainder of the deliverable is structured as follows: chapter 2 discusses the SPEED-5G use cases. Chapter 3 provides enhanced functional and system architecture, chapter 4 discusses on degrees of virtualisation, chapter 5 discusses performance metrics, and finally chapter 6 elaborates on the conclusions of this deliverable.

# 2    SPEED-5G scenarios and use cases

The main objective of SPEED-5G is to develop novel solutions to address 5G technology trends and requirements. Specifically, the main scenarios investigated in SPEED-5G refer to indoor and indoor/outdoor scenarios (around buildings) where capacity demands are the highest. SPEED-5G considered the following four eDSA scenarios, which have been discussed in D3.1:

- Massive IoT communications (mIoT)
- enhanced Mobile Broadband (eMBB)
- Ultra-Reliable Communications (URC)
- High-Speed mobility

These scenarios covered the SPEED-5G concepts for urban communications, IoT, mission-critical, and vehicular services and as a result are representative for validating RRM/MAC solutions and eDSA techniques that SPEED-5G proposes. Moreover, these eDSA scenarios are further divided into three realistic use cases which are described in section 2.1 and will be demonstrated in WP6.

eDSA is an innovation of SPEED-5G, the aim of which is to increase the efficiency in the use of radio resources, especially spectrum but also transmit energy, while providing the high levels of user experience which have been defined as use cases and KPIs in section 5.4. SPEED-5G works with spectrum below 6GHz and it is in these bands that harmonised spectrum is in particularly short supply. Pressure on the spectrum below 6GHz is expected to further increase with time, so that systems must be increasingly smart about how they use it.

eDSA consists of two major parts, which are the Radio Resource Manager (RRM) and the MAC layer of the base stations. The RRM is further explained in WP4 and MAC in WP5. SPEED-5G high level concept of eDSA and the mapping to work packages is shown in Figure 1. The output from WP3 goes to WP4 and WP5 to design and model RRM and MAC tailored to the use cases. Moreover, the definition of new RRM/RRM entities, functions and interfaces for each of these use cases will be defined in WP4 and WP5. Finally WP6 will deploy and demonstrate the selected use cases in the testbed.

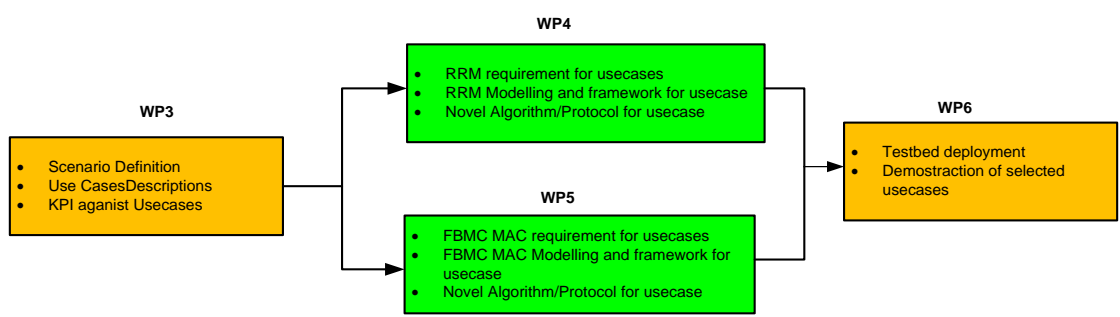

*Figure 1: Interaction between WPs*

## 2.1 SPEED-5G use cases (UC) description

The RRM-MAC functionality is shown in Figure 3, which is high level and is designed to cope with all three use cases to be demonstrated. These use cases are dynamic channel selection, load balancing and capacity augmentation in small cells. Moreover, it is necessary to refine the function processes for theses use cases, which will be detailed in WP4 and WP5.

### 2.1.1 Dynamic channel selection (UC1)

An initial situation is envisaged where multiple small cells are managed by a single network operator. The small cells are conveying a mix of traffic that corresponds to the traffic in the SPEED-5G use cases: enhanced mobile broadband (eMBB), ultra-reliable communications (URC), massive IoT (mIoT) and high SPEED mobility. There would be an assumed amount of interference that could be varying, degrading the QoS.

A decision is made by the RRM to select a band from a combination of licensed and unlicensed spectrum available and possibly also to choose a different RAT, depending on the context and the ability of the user terminal. The aims would be

- to meet some specified QoS requirement
- to allow for cell selection as one mechanism
- to allow for opportunistic sharing as an option, similar to TV whitespace
- to be backwards-compatible with existing LAA and LTE-U standards.

The decision as to which channel (i.e. sub-band) within the band is then delegated to the cell, which, for example, can use the exponentially-weighted randomised algorithm (WRA). WRA and refinement of function processes for the case of dynamic channel selection will be explained in WP4 and WP5.

### 2.1.2 Load balancing among a group of neighbouring small cells, to manage interference using non-licensed spectrum (UC2)

Let us consider a group of neighbouring small cells managed by the same operator and sharing the same licensed spectrum. These small cells are conveying a mix of traffic composed of broadband, ultra-reliable communications and IoT. Due to the high traffic load, the dense deployment of small cells and the overlay of the macro cell, which may operate on the same channel, the level of interference may be too high. This leads to an excessive degradation of the QoS/QoE.

A centralised RRM controller (located in the edge virtualised architecture) analyses the load on each small cell, the level of interference on the licensed band (e.g. by means of CQI reports, BLER measurements, and/or NACKs). If an acceptability threshold is down-crossed, the controller decides to trigger a load balancing process. Based on the characteristics of the different available shared bands (bandwidth, central frequency, path loss and regulations), the default mapping of traffic types on spectrum resources (depending on the license regime, available bandwidth and regulation limitations) and the estimation of the interference of these bands, part of the data traffic is steered to some non-licensed bands.

In particular, the traffic steering function is in charge to select the traffic types that can be handed-over on a specific band. Furthermore, the RAT/spectrum selection identifies, for each band, the maximum bandwidth, suitable transmit power, RAT and default MAC frame configuration, depending on the regulation constraints (e.g. listen-before-talk maximum period) and the interference level. The control traffic is kept on the licensed spectrum.

Regarding the offloaded traffic, at the MAC layer, a suitable channel is selected for each traffic type and/or for each band. This can be done either by relying on prior available sensing measurements already available or by resorting on the sensing capabilities of the small cells. Additionally, a MAC frame format is configured, using either a default configuration provided by the RRM part or through an adaptive MAC frame configuration.

For bands requiring a listen-before-talk procedure, the MAC layer trigger for initiating the actual transmission of a frame is provided by the CCA function. This functional block receives the PHY measurements about channel occupancy (e.g. based on the energy level) and decides whether the channel is available and can be accessed. If the channel is available, the MAC triggers the scheduler to initiate the process of mapping resources onto the frame format, allocating uplink and downlink traffics in physical resource blocks (time and frequency resources). Additionally, based on the PHY measurements (specifically the noise level in the channel), this block may estimate a CQI value that can be provided to the scheduler as an additional updated information about the channel quality. Refinement of function processes for this use case will be explained in WP4 and WP5.

### 2.1.3 Small cell throughput improvement with carrier aggregation (i.e. offload part of the traffic) to non-licensed spectrum (UC3)

Recently, wireless local-area network (WLAN) technologies based on the IEEE 802.11 standards (e.g. Wi-Fi, Wi-Fi Direct) and wireless personal-area network (WPAN) technologies (e.g. Bluetooth, Ultra-Wideband [UWB] technologies) have been increasingly used, due to proliferation of smartphones and increasing number of people that now live in cities. These technologies are designed for short distances between sender and receiver and therefore achieve very high data rates with low energy consumption. However, communications on a licensed band of a cellular network can be better in terms of interference avoidance under a controlled environment. The following are shortcomings of the above-mentioned technologies.

As Wi-Fi and Bluetooth work in licence exempt bands, there are no guarantees that they work everywhere since there is always the possibility of the presence of an interfering communication system or other sources of interference. Wi-Fi Direct can be used in every public place in the near future as devices become available, but this technology lacks global synchronisation can be used in wireless systems, generally to enable energy-efficient operations. For devices to discover each other, they must rendezvous in space and time. Only in a synchronised system the discovery periods can be both frequent and of low duty cycle. Thus, in practice, devices operating autonomously without infrastructure support in unlicensed spectrum can synchronise, but only locally.

In order to solve the above-mentioned issues, recently the device-to-device communication (D2D) concept has been proposed by LTE-A. D2D allow users to communicate directly (send data directly on licensed/unlicensed band) without access to fixed infrastructure under control of operator (licensed band). The potential advantages of D2D communication are the offloading of traffic, throughput enhancement, coverage expansion, and UE energy saving.

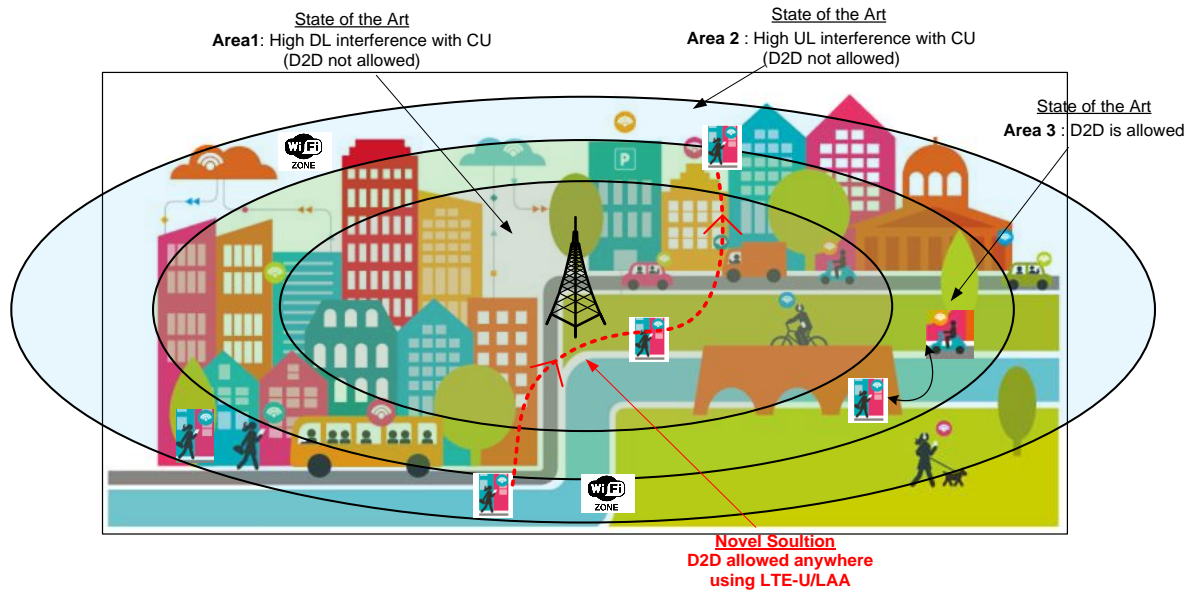

*Figure 2: D2D offloading on LTE-U/LAA*

In a 5G network, D2D coexists as another tier with the small cells network, which can operate in either licensed/unlicensed band. A D2D link will reuse the cellular resource, which create two types of interference: 1) Intra-cell cross-tier interference between the D2D and cellular users (CU) and 2) Inter-cell interference between the D2D links in coverage area of different BS.

LTE release 12 and state-of-the-art literature on D2D suggest using the LTE uplink band for D2D communication (this performs much better than D2D sharing the DL). However, D2D sharing the UL band leads to a higher interference to normal CUs. Therefore, there is trade-off between D2D and CU performance while considering whether to use UL or DL band. Letting D2D transmission utilise the DL band, favours CU reliability over D2D reliability whereas, letting D2D transmission utilise the UL band will favour D2D reliability over CU reliability [17].

Moreover, geometric areas (see Figure 2) are also important to the performance trade-off between D2D and CU (to use UL or DL band).

- Cell centre (area 1) is generally off-limits to D2D transmission using DL band
- Cell edge (area 2) is generally off-limits to D2D using UL band
- If only cellular DL/UL bands can be used, reliable D2D communication would be kept away from cell centre/edge.

Therefore, to solve the above problem in D2D communication, this use case considers the D2D communication using lightly licensed, unlicensed and TV white spaces using LTE-U/LAA-like mechanism. By using LTE-U/LAA, D2D can operate anywhere in the cell coverage, except for the region where other unlicensed band RATs are in use. D2D communication in licensed/unlicensed bands under operator control offloads the traffic, possibly using single/different RATs. Traffic with the most stringent QoS requirements is allocated on licensed or lightly licensed spectrum using normal cellular operation. Traffic with non-stringent PER and latency requirements can be moved on D2D using licensed/unlicensed bands. Predetermined association rules (traffic class vs. spectrum band) could be established: i.e. high data rate (QoS: 2.6 GHz, Non-QoS: 5GHz, 2.3 GHz (ASA), 700MHz (TV white spaces)).Refinement of function processes for this use case will be explained in WP4 and WP5.

# 3 Architecture for supporting eDSA

The SPEED-5G architecture design principles are based on flexibility, simplicity, on-demand resource allocation, auto-scaling and enhanced performance..

Figure 3 provides a high level view of the blocks which are considered in RRM and MAC and are described thoroughly in WP4 and WP5 respectively.

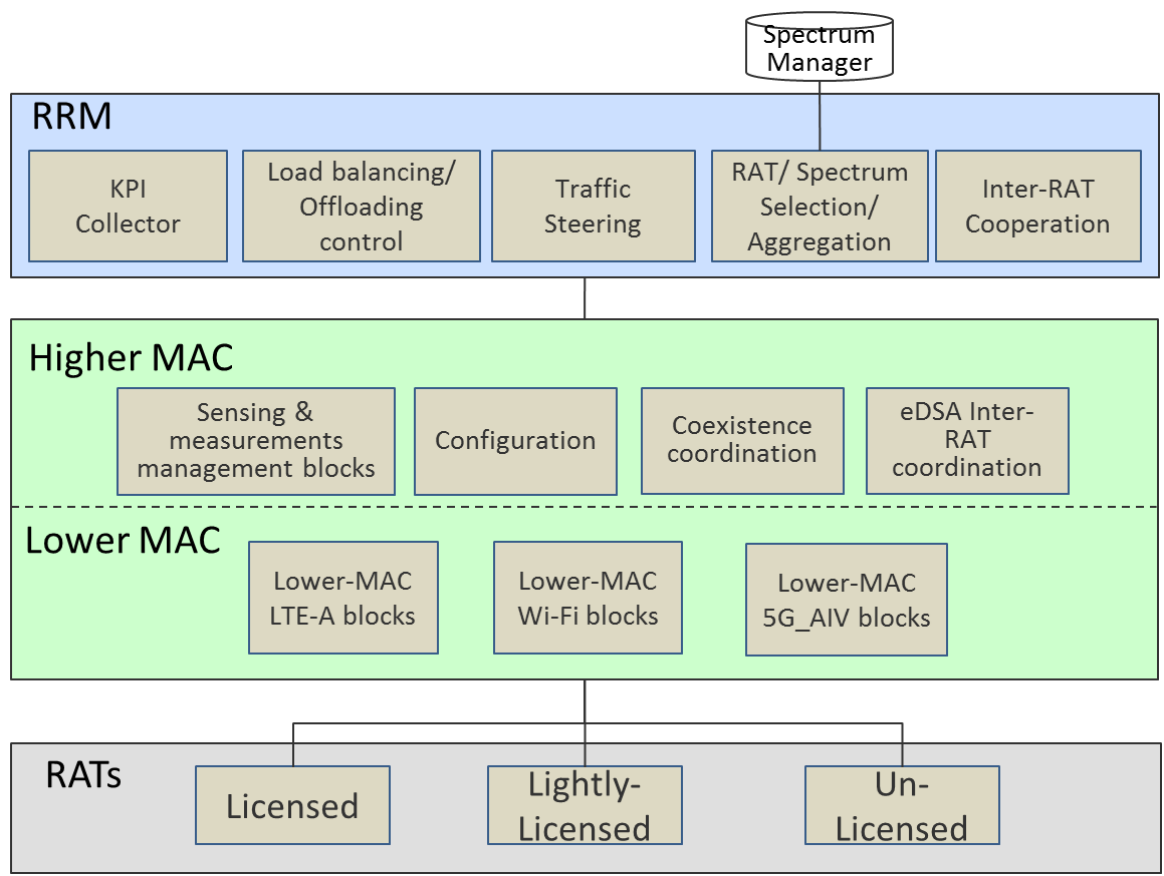

*Figure 3: Architecture for supporting eDSA*

A link that provides feedback to the RRM related to channel monitoring is provided. The purpose of this is that with licensed spectrum, the interference level will affect the RRM decision on the channel and channel aggregation patterns. The MAC layer is responsible for allocating resource blocks or their equivalent in any new 5G RATs, and this can include ICIC constraints where cooperation is needed between neighbouring base stations on co-channel. These functions are typically managed using distributed SON functions, which aim to improve the performance of cell-edge users. For unlicensed spectrum, the RRM will specify the frequency band, such as 2.4 or 5GHz, and it may also need to steer the MAC layer to a channel in certain situations, despite the earlier statement that the MAC is free to choose the unlicensed channel. For example, if lightly-licensed bands can be used from multiple base stations, a degree of coordination will be needed. The eDSA architecture may well develop further as the project progresses, for example mapping onto a practical network architecture that involves fronthaul and backhaul.

## 3.1.1 RRM-related blocks

**Load balancing/Offloading control**

Load balancing aims at making efficient use of the limited spectrum to deal with unequal loads in order to improve network reliability by reducing the congestion probability in hot spot areas of cellular networks. Since the small cells envisioned by SPEED-5G are able to handle and transmit traffic using different RATs and different spectrum bands, this functional block in SPEED-5G adds the capability to trigger the offload of traffic to unlicensed/lightly-licensed bands or to less loaded bands.

> SPEED-5G proposes novel algorithms which can deal with problematic situations in a proactive manner by utilizing machine-learning principles which will lead to the realization of intelligent nodes. Specifically, through machine-learning it will be possible to learn the most appropriate band (licensed/unlicensed/lightly-licensed) according to the service that we want to handle and proceed to the assignment.

### Traffic steering

Traffic Steering is a key functional block of the SPEED-5G RRM. Its main objective is to provide a labelling of the different types of traffic and a mapping between types of traffic and the available bands. This functional block decides what traffic types can be assigned to what available bands. Some pre-determined association rules could be established, depending on the requirements of each specific traffic type and depending on what spectrum bands or RATs are available. For instance, traffic with the most stringent QoS requirements may be allocated only on licensed or lightly-licensed spectrum, which provide ways to ensure the provision of some end-to-end quality indicators. On the other hand, traffic with non-stringent error and latency requirements may be moved to unlicensed bands.

> SPEED-5G proposes novel machine-learning algorithms that can learn what are the most suitable bands for certain types of traffic and assign them accordingly as soon as the expected traffic is observed.

### RAT/Spectrum selection and aggregation

The objective of this functional block is to select a suitable band and RAT to be used by each type of traffic. It also selects the number of channels to be used within a band, if needed. This functional block determines the system spectral efficiency and therefore how much aggregation is needed.

> SPEED-5G proposes algorithms for dynamic channel assignment and prioritization of traffic according to certain criteria. Criteria may be the user class (e.g., gold, silver, bronze users) or the type of service (e.g., critical service, non-critical mobile broadband service etc.)

### Inter-RAT cooperation

Some new 5G and legacy RATs will coexist in the 5G network. The cooperation of these different wireless technologies improves the network operating efficiency and user experience. This block aims at improving the coexistence with other RATs in the same band, e.g. in the 5GHz unlicensed band where the data transmissions must coexist with WiFi.

> This cooperation considers the following aspects:
>
> - Intelligent access control and management: according to the network state, wireless environment, UE capabilities, the cooperative sensing and the eDSA paradigm, each service will be mapped to the most appropriate RAT, improving the user experience and network efficiency.
>
> - Multi-RAT wireless resource management: according to service types, network loads, interference levels etc., RRM is expected to support multi-RAT joint radio resource management and joint optimisation, which realises the interference coordination among multiple RATs, achieving resource sharing and allocation.

**Spectrum Manager**

The Local Spectrum Manager entity coordinates the eDSA function for different clusters of cells, allowing taking spectrum allocation decisions over the unlicensed and lightly-licensed spectrum utilization. This network entity also includes local/ regional cognitive radio databases and a set of tools for enabling 5G systems to operate under different spectrum sharing scenarios.

**KPI Collector**

The KPI Collector is responsible to collect the set of KPIs from each cell in a centralized way. Current standards define an xml format to report the KPIs from the cell to a centralized entity. The communication protocol is not standardized but no real time is required, neither a permanent connection between the KPI Collector and the cells.

## 3.1.2    MAC-related blocks

SPEED-5G proposes the decomposition of MAC layer into 2 sub-layers: high MAC and low MAC. High MAC is composed of a set of functions which are RAT-independent and meant to coordinate the underlying RATs. It covers the coexistence management, transmission opportunity identification, the implementation of real time functions of traffic steering, load balancing, scheduler configuration, multiple access coordination, to name a few.

At the contrary, low MAC functions are RAT-dependant functions, mainly related to scheduling, logical channel management, bearer configuration, channel (de)multiplexing and (de)framing. In some way, high MAC is seen as a convergence point of the protocol stack dealing with the control path, decoupled from the user plane and managing the sets of possible bearers.

**Sensing & measurements management blocks**

This group of functions is responsible for collecting sensing measurements and link control KPIs and forwarding them to RRM. It includes "Measurement Reports" and "Sensing Results".

**Configuration**

Configuration parameters related to the cell, the UE and the scheduler are important for SPEED-5G use cases and scenarios. Reconfiguration is done by the RRM algorithms in cell, UE and scheduler that are checking the reported KPIs. The configuration data has to be easily accessible and maintain the integrity for scalability purposes.

**eDSA Inter-RAT coordination and coexistence coordination**

These are the core set of functions which enable the eDSA at MAC level which include functions used to manage traffic steering, logical channel management for RAT and spectrum aggregation, multiple access, frame formatting and broadcast. This group of functions are actually applying coarse grained decisions taken at RRM level, taking into consideration real time conditions experienced at lower MAC level.

SPEED-5G brings the following advancements in MAC:

- Ability to address both an improvement of existing technologies based on carrier aggregation using bands under diverse license regimes and a disruptive distributed TDD MAC protocol where multi-connectivity and spectrum aggregation are built-in features.

- Native support for multi-RAT operation as 5G networks are expected to unify a broad set of RAT under the same network architecture.

- Decomposition of the MAC layer into a high MAC layer handling RAT cooperation and logical channel management and a low MAC supporting the RAT-specific functions.

# 4    Degrees of virtualisation and influence on access

## 4.1    Small cell virtual deployment

The SPEED-5G project architecture is based on Software-Defined Networking (SDN) and Network Function Virtualisation (NFV) while the Small Cell (SC) is based on the Software-Defined Radio (SDR) concepts. The main goal of the SDR concept is to dynamically provide the most appropriate radio network functionality where virtualisation allows that some specific hardware functions are implemented through software. The virtualisation provides flexibility by assuming extra computational process. The SPEED-5G architecture is ready to support virtualised and non-virtualised stacks as can be, for instance, a cloud-RAN solution for a virtualised solution or the stack deployed in a femtocell. A femtocell is a low cost small cell designed to cover short range indoor areas which due to its HW capabilities and final usage, virtualisation procedures are not recommended.

The virtualisation facilities are provided by NFV and SDN concepts. NFV provides the network abstraction between the physical hardware (HW) and its abstraction mechanism. In that sense, a physical layer may be abstracted to logical networks. The SDN concept is related with the software (SW) and its main characteristics is that it decouples the control-plane from the data-plane. Figure 4 shows the basic SDN architecture where the Application Layer contains several network functionalities as the centralised Radio Resource Management (cRRM). In addition to the protocol stack, it is important to consider that an operator's network is composed by many other entities. Therefore, entities as the HeNB Management System (HeMS), Cores or the Auto-Configuration Server (ACS) have to be also supported. The application layer functionalities run over the Infrastructure Layer where the real equipment is deployed. For that reason, all the functionalities included into the application layer are instantiated in different networks elements.

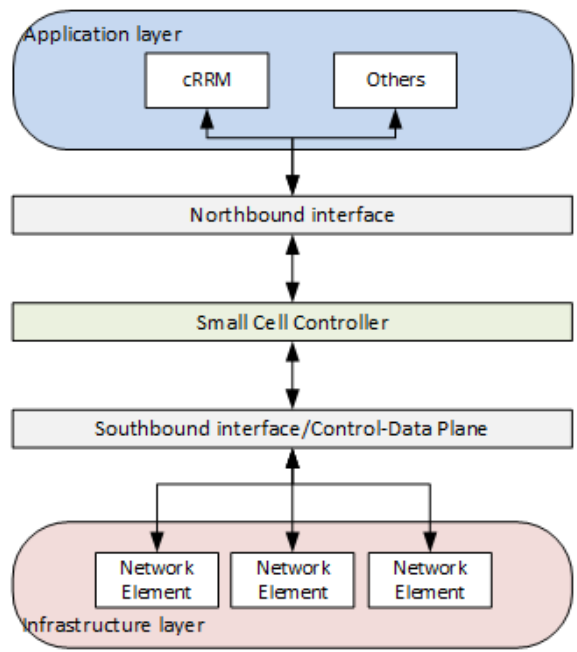

*Figure 4:  Basic SDN/SDR architecture*

Figure 5 shows the protocol stack proposed by SPEED-5G where a cRRM configures the stack layers, mainly referring to the Medium Access Control (MAC) layer. More details about the RRM SPEED-5G procedures for (re)configuring the stack layer are defined at WP4 in D4.1. In the same way, how the MAC layer is (re)configured is defined in detail at WP5 in D5.1. The required interfaces and its communication between the RRM and the MAC layer is out of the scope of this document.

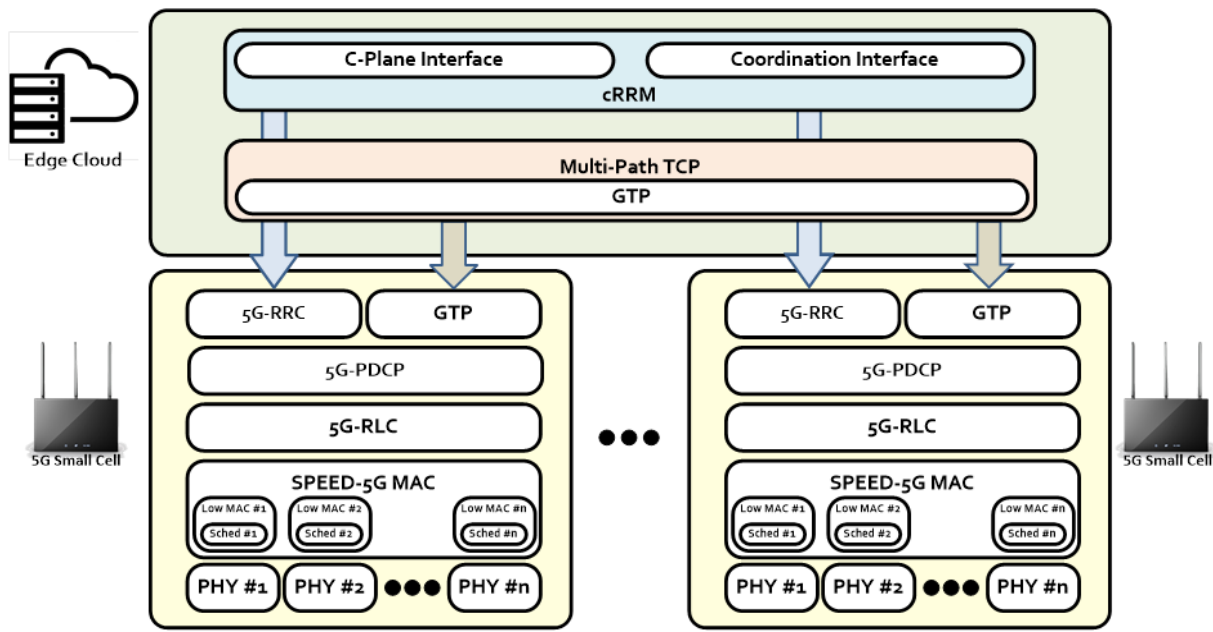

*Figure 5: Protocol stack proposed by SPPED-5G*

The proposed protocol stack in Figure 5 is designed to support network slicing. A network slice is a novel concept related with the end-to-end user requirements. The main idea behind this concept is to dynamically provide the most optimal network in order to achieve a specific goal. The network slice is an accepted concept but is still pending to define what an optimal network slice is since there are multiple parameters that are taken into account. For instance, these two different visions about what an optimal network is. One vision is to provide a specific slice per service, that is, the virtual network is optimised to provide broadband or ultra-reliable services in a specific area. Another vision of what is an optimal slice is relates to device characteristics. A specific slice may then be optimised for frequent handovers or Device-To-Device (D2D) communications. That is, without considering the specific user service. Although as yet there is no consensus among vendors and operators as to how best slicing should be done and how slices are to be managed, the proposed architecture is flexible enough to support different approaches.

For a better understanding, two different deployment examples can be considered using the same use case: i) configure/instantiate the MAC layer depending on the Load Balancing RRM requirements and ii) virtualization The two examples considered (detailed below) show how the virtual layers are instantiated and, when required, the MAC layer is configured (in example 1) or instantiated (in example 2). Both examples make use of the *OpenStack [2]* , an open-source orchestrator used in cloud-computing. The first example describes the SPEED-5G proposal where the RRM is centralised and the protocol stack is deployed into the SC. This solution is aligned with split 2 of Figure 7 where the different stack splits proposed for LTE are shown. In the first example, the centralised functionalities are virtualised while the functionalities at the SC are not, whilst the second example assumes that there are no technological restrictions on the fronthaul and everything may be virtualised, as per split 6 in Figure 7.

The OpenStack orchestrator functional architecture is depicted in Figure 6. OpenStack is composed by many different functionalities and only the most relevant blocks are described next. OpenStack functions are divided into the computing service called Nova; the networking named Neutron; the image service, Glance; the block storage, Cinder and finally; the use case and trigger analyser, the Telemetry. Nova is the network brain which receives triggers/alarms. Telemetry is constantly monitoring the desired parameters (as CPU or memory usage) of the virtual network ensuring that it is working in an efficient way. Thus, the triggers/alarms are periodically evaluated and, when required, the network reconfiguration starts. It is Telemetry that raises the alarm/trigger to Nova which, depending on the input trigger/alarm, requests *Glance* to instantiate the appropriate image into the network. *Nova* also asks *Neutron* to create all the required network management and connections between the current instances and the new one. Finally, *Nova* requests *Cinder* for the appropriate data storage for instance particular instance ensuring that new instances can work without any issue. More details about OpenStack can be found at [2].

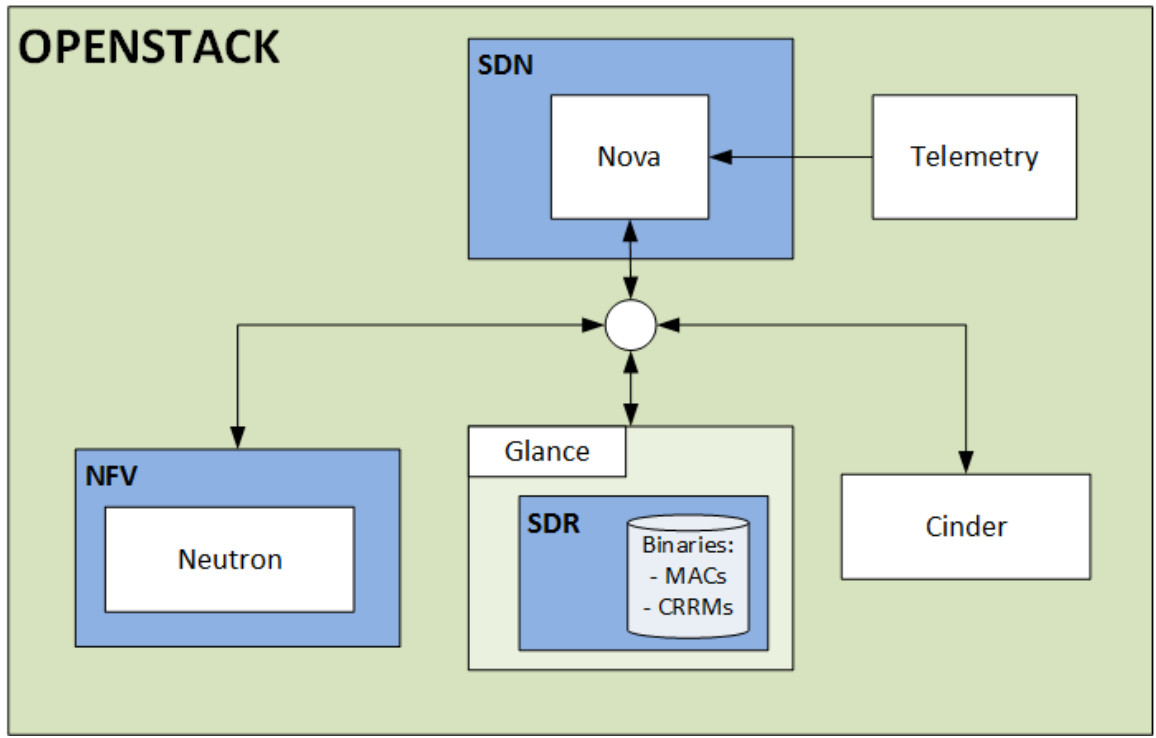

*Figure 6: Map between procedures and OpenStack functionalities*

After the introduction of OpenStack functionalities, we can now proceed to the description of the examples. The first example is concerned with intial cRRM instantiation, how the protocol layer is installed into the SC and finally, how the cRRM reconfigures the MAC layer. The steps are the following:

1. The process starts. It is assumed that *OpenStack* framework is running and only cRRM and the protocol stack is pending to be instantiated or installed

2. OpenStack instantiates cRRM

   2.1. *Nova* asks *Glance* for the cRRM entity and *Cinder* to allocate the appropriate resources

   2.2. *Neutron* is called by *Nova* in order to create the network connections into the system

   2.3. The cRRM is instantiated and *Telemetry* checks the physical equipment resources in order to instantiate a new one or remove it when required

   2.4. cRRM is configured to execute its appropriated algorithms depending on the SC characteristics

3. The MAC layer has to be selected

   3.1. *Nova* asks *Glance* for the most suitable MAC from a set of them, depending on the physical cell characteristics as, for instance, its available radio interfaces

   3.2. The selected protocol stack, containing the previous selected MAC layer, is installed during the bootstrap and the SC is started

4. At this point everything is working and connections are established. Thus, the cRRM is constantly monitoring the spectrum data and the KPIs until the load balancing algorithm raises a trigger

   4.1. After a trigger, the cRRM decides the new MAC configuration including all the required information to properly execute the load balancing

   4.2. New configuration is sent and the MAC layer is reconfigured

5. MAC layer is reconfigured in order to achieve the requested procedures. Notice that this MAC layer already supports these functionalities, nothing related with virtualisation

In the second example, the final solution is based on Cloud-RAN. The example is addressed to show how the cRRM and the MAC layers are instantiated and, when required, the cRRM raises the requirement to instantiate a new MAC layer.

1. The process starts. It is assumed that *OpenStack* framework is running and only cRRM and the protocol stack is pending to be instantiated or installed.

2. OpenStack instantiates the cRRM module and the SC protocol stack

   2.1. *Nova* requests *Glance* for the cRRM and the MAC binaries to be instantiated. In this case, in order to show the Cloud-RAN capabilities of our solution, the initial MAC of the instantiated SC stack does not support the procedures that cRRM will reconfigure

   2.2. *Nova* requests *Cinder* for cRRM and protocol stack memory space

   2.3. *Neutron* is requested by *Nova* to create the network connections

   2.4. cRRM and the MAC layer are instantiated. After that, *Telemetry* starts to check if the assigned computational resources are enough and, in case that they are not, another cRRM or MAC have to be instantiated. In the same way, when one instance is not required any more, it has to be removed

3. cRRM triggers the requirement to use LAA due to load balancing algorithms

4. Step 2 and its sub points are repeated but only for the MAC and, when all the steps are done, the older MAC will be removed

The examples discussed above provide an introduction to how the system works using virtualization procedures for centralized function. The main idea is to ensure incorporation of the relevant SPEED-5G entities within a virtualized framework that natively supports SDN and NFV functions.

## 4.2    Virtualisation split at the different network sub-systems

Focusing on the protocol stack, when the stack has to be virtualised, the application layer contains two different applications, one for the centralised stack and the other one for the distributed stack functions. This functionality split is managed by SDR functions.

The radio protocol stack can be virtualised where some functions are executed in the cell and other functions are executed into the cloud. There are several possible splits that depends on the stack. Each one of these splits has advantages and also disadvantages. Every one of them has to be analysed in order to have all the information to take a decision.

The following section focuses on the different splits considering only the LTE stack. The overall LTE description may be found at [3] where interfaces, network architecture including the description of its elements, mobility management and the layers of the protocol stack and its functionalities are explained. If another RAT is required, the same study has to be done.

## 4.3 Stack virtualisation based on 3GPP model

This section focus the stack virtualisation based on the LTE stack. Assuming this restriction, Figure 7 shows different protocol stack splits where, from a technical point of view and without considering any constraint, centralized entities are always beneficial or, at least, they have to be able to work in the same way than if they are distributed. For that reason, on each split, the advantages will be identified. On the other hand, a system like LTE has a big temporal constraint since the Time Transmission Interval (TTI) is 1 millisecond. For that reason, the roundtrip time between distributed and centralised operations has to remain within this temporal restriction. Therefore, the main network restrictions in a centralised split is the fronthaul bandwidth (BW) and its latencies. The study is done over LTE by ETSI, which has defined several use cases [3].

The network virtualisation split has been fully studied and the common consensus is shown in Figure 7.

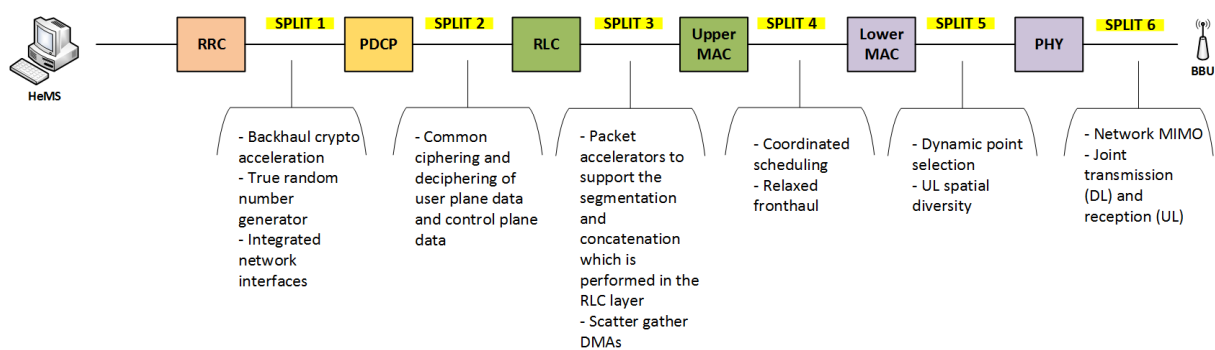

*Figure 7: Virtualisation splits [3]*

The LTE stack is divided into different layers who are well defined by 3GPP series 36:

- layer 3 is dedicated to control functionalities. Layer 3 is composed by the RRM and the Radio Resource Control (RRC)

- layer 2 includes compression, security and scheduling. It includes the Packet Data Convergence Protocol (PDCP), the Radio Link Control (RLC) and the Medium Access Control (MAC)

- layer 1 regards the physical layer.

For the LTE protocol stack, 6 different splits are possible. Split 1 virtualises Layer 3. From the split 2 to split 5 the L3 and the L2 is virtualized. The Split 6 virtualises the complete stack.

**Split1: RRC**
The RRC layer protocol is virtualised into a data centre. The virtualised functions provided by RRC consist of application services, management and RRC functions. The control plane stack in the centralised cell supports control data interface towards the core network and the relay of RRC control messages towards and from the remote cells over a control. The user plane traffic is sent through the GTP to the PDCP layer. The GTP entity has to be also centralised at same level as the RRC layer.

The main advantages of a centralised virtual RRC are:

- Backhaul crypto acceleration

- True random number generators

- Integrated network interfaces

- Cell and radio configuration
- Centralised connection mobility control, measurement reporting and handover trigger control

Most of the stack functionalities are done in the own cell. For this reason, the fronthaul requirements, including bandwidth and delays, are feasible.

**Split 2: PDCP**

This split proposes virtualisation at PDCP layer and above, and has some commonality with the 3GPP Rel. 12 study [3] that outlines the usage and architectural options for dual connectivity. From every layer, the identified advantages are added to the previous identified ones. One of the main function of PDCP layer is the user plane ciphering so, in that direction, the main advantage is:

- Common ciphering and deciphering of user plane data and control plane data

This allows ciphering to be maintained even when the device moves from one cell to another.

**Split 3: RLC**

This split proposed virtualisation at RLC layer and above. Since RLC contains the UE buffers of each traffic type, the fronthaul between the RLC and MAC layer requires a very low latency. Each traffic type is mapped into a specific logical channel creating a univocal relation between traffic type and logical channel. For that reason, from this point forth, talking about logical channels is as talking about traffic types.

The QoS is offered by prioritizing specific logical channels but this task is done at the MAC layer. It is important to notice that, in a LTE system, the MAC layer notifies the RLC layer the amount of data per logical channel and this is the reason to have a very low latency fronthaul between RLC and MAC layers.

The RLC virtualisation offers mobility enhancement when it works in Acknowledge Mode (AM). Working in AM, the retransmission are done at this level then, DL RLC retransmission does not require extra messages between the source and the target cells.

The main advantages at this split are:

- Reduce processor requirements on the cell since the periodic status PDUs(de) is virtualized
- Mobility seamless is AM mode

**Split 4: Upper MAC**

The upper MAC layer includes one of the main functionality, the scheduler. The centralised upper MAC works with the associated RRM and OAM who are used to configure the layer. A centralised scheduler provides multiple benefits allowing joint-scheduling or Multi-User Multiple-Input Multiple-Output (MU-MIMO). The SPEED-5G protocol stack is ready to support Multi-Operator Core Network (MOCON) where more than one operator sends traffic to the same cell. In a network where MOCON is supported, a centralised entity knows the cell or cells where the device is connected and, consequently, is able to perform the scheduling process in an optimal way.

The scheduler works with the Hybrid Automatic Repeat Request (HARQ) entity responsible to manage retransmissions. Note that the HARQ entity only manages the retransmissions but the PDU packets are stored into the physical layer. Since in this split the physical layer is decentralized, a fully mobility seamless cannot be achieved.

The main advantages provided by this split are:

- Coordinated scheduling
- Relaxed fronthaul because all the messages between RLC and MAC layers are removed from this interface

**Split 5: Lower MAC**

The lower MAC layer functions include the Channel State Information (CSI) process. This information is sent by the device to the cell. Assuming two or more cooperating cells and the centralised functionality, it is possible to optimise Coordinated Multipoint (CoMP) techniques.

Two other important functionalities of the MAC layer are (de)multiplexing and (dis)assembling. When a device has dual connectivity, the data is sent to more than one cell. Then, when the centralised MAC layer receives both contributions, it is able to recompose the message send by the device.

The main advantages of this split are:

- Dynamic point selection
- UL spatial diversity

**Split 6: PHY**

The centralised small cell function includes a Base Band Unit (BBU) that represents components (hardware and software) used to perform L1 PHY processing. This split allows to consider the antennas of the network as a massive MIMO with all its well-known features. The BW fronthaul for supporting this split is around 5Gbps. Consequently, this split will be feasible in the near future.

The advantages to have a complete virtualised stack are:

- Network MIMO
- Joint transmission (DL) and reception (UL)

**Fronthaul Requirements**

Each split adds the previous advantages plus the own split advantages. In that sense, the more the stack is virtualized, the more benefits can be obtained and therefore, split 6 is always the most convenient.

The problem is that this is true only when the fronthaul is ideal in terms of bandwidth and latency. Since practically this not always true, the fronthaul BW and latency requirements for each split are required. The results of study in [22][23] is summarised in Table 1 where the fronthaul requirements in terms of BW and latency are shown for each split.

| Splits | Bandwidth and latency | |
|---|---|---|
| | BW | Latency |
| Split 1: RRC | 200Mbps | < 6ms |
| Split 2: PDCP | 200Mbps | < 6ms |
| Split 3: RLC | 200Mbps | < 6ms |
| Split 4: Upper MAC | 200Mbps | 250us – 2ms |
| Split 5: Lower MAC | 625Mbps – 2Gps | 250us – 2ms |
| Split 6: PHY | 5Gps | 250us |

*Table 1: Study of required fronthaul bandwidth and latency [22]*

The conclusion is that, based on the above table, the maximum allowed split depends on the fronthaul constraints of bandwidth/capacity and delay.

## 4.4    Degrees of virtualisation and the effect of fronthaul

By degrees of virtualisation here we mean the degree to which base stations have their functions virtualised within a local/ regional datacentre or server facility. Figure 8 shows a high level virtualisation concept.

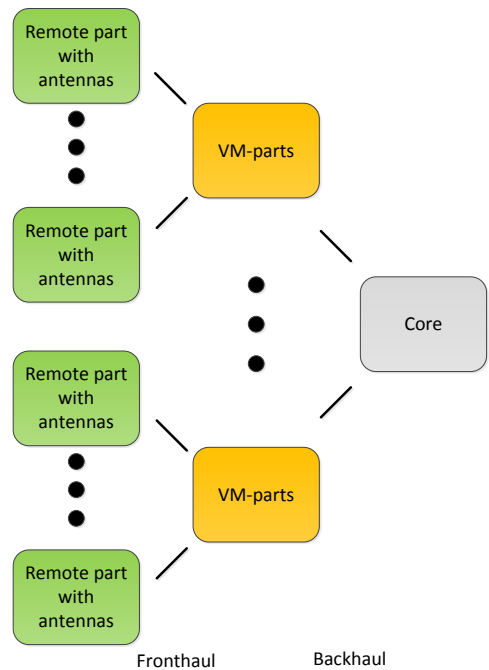

*Figure 8: Virtualisation high-level parts*

In the figure, the remote parts with antennas are installed in the physical locations where the antennas are needed, for example inside houses, or upon masts. The virtual machine (VM) parts are located centrally or semi-centrally, for example at aggregation points where they may connect to several hundred or thousand remote parts using the fronthaul. The VM parts are in turn connected to a core using the backhaul. In a typical national network, there may be one or two cores, with resilience, several thousand VM-parts that may be located in buildings such as telephone exchanges, and several million remote parts in the customer premises or up masts.

There is benefit in putting as much functionality as possible into the VM-parts, and the benefit derives from two sources. Firstly, the software in the VM-parts can be run on servers that are low-cost and non-special, and they are owned by the network operator. Secondly, the system can be more often upgraded by upgrading the software in the VM-parts without replacing the remote parts hardware, which reduces the cost.

The functions that are on either side of the fronthaul can be varied. In theory, the minimum that can be put into the remote part is just the antennas, with everything else in the VM-parts, and this means that the RF signals pass over the fronthaul by means of RF cables or RF over fibre. In practice though, the minimum that can be placed at the remote part is the antennas plus RF components like the transmitter power amplifier, receiver LNA, some filtering and the up and down converters. This means that the base-band IQ samples must be transported across the fronthaul, which would result in several hundred Mbit/s and probably higher if MIMO is used. The CPRI standard interface is designed for this purpose and commonly it needs fibre connection. The maximum that can realistically be put into the remote parts is the RLC, MAC and PHY functions, which means that just the packets flow over the fronthaul, and the fronthaul can be a lower-bandwidth technology such as wireless or copper. There is a trade-off in deciding how much functionality to put into the remote part: the less put in the remote part the more is virtualised and more cost savings can result. On the other hand, this means that the fronthaul has to carry a higher bit-rate and possess a lower latency. Thus there is a trade-off in the cost savings from virtualisation against the increased costs of providing the required fronthaul capacity and low latency. This poses the problem of where to split the cells into remote parts and VM parts and this problem is investigated in this section.

With LTE, when the UE tries to attach, it requires a response from the MME within 8ms otherwise it re-tries, and this puts a latency constraint on the system including any fronthaul. Another difficulty that arises with virtualisation is the ability to use LIPA (Local IP Access), due to the need to route the IP packets from the LIPA gateway that is now in the server centre, back to the local network.

### 4.4.1    Cell splitting options

Figure 9 shows a simplified SPEED-5G base station protocol stack, where multiple PHY entities are shown that are scheduled by the MAC. Each PHY entity generates baseband signals that are forwarded to the RRH or possibly more than one RRH even though only one is shown in the diagram.

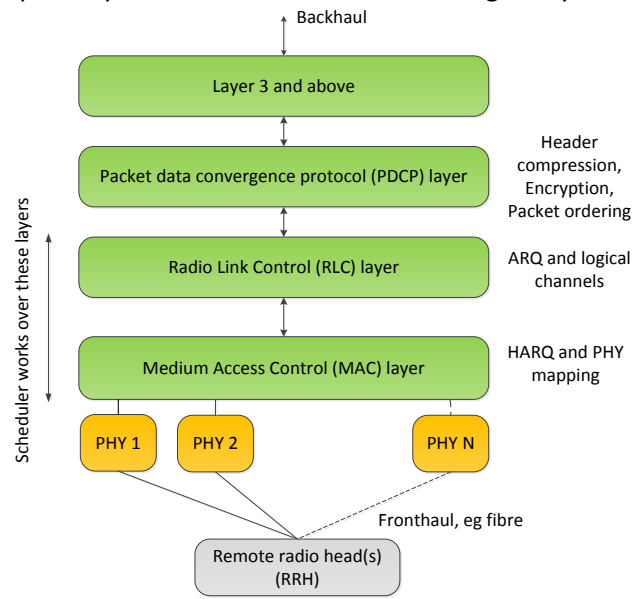

*Figure 9: Putting the minimum at the remote parts – the antenna and RF components – means that the fronthaul has to carry very high bandwidth*

The scheduler works over three layers of the stack, the RLC, MAC and PHY layers. Therefore, in this case the scheduling is done entirely in the VM part, which has the benefit that the scheduler performance is not affected by the fronthaul. On the right-hand side of Figure 9 we show the basic functions of the layers. Putting the fronthaul as shown in the figure does not require any innovation; it is covered by the CPRI standard. It does however depend on the presence of fibre and preferably dark fibre, which is not generally installed in domestic premises in Europe and nor it is generally available even to connect macro cells[2]. Putting the fronthaul as shown in the figure also poses problems for the LTE uplink, since the HARQ is synchronous and requires acknowledgements 4ms later in the frame. If the latency on the fronthaul is too high to allow this, the uplink packets must be falsely acknowledged and the link budget must carry a higher safety margin to prevent significant errors and re-transmissions. Typically, this will mean that a lower modulation rate is used on the uplink and hence a lower uplink bandwidth will be available. To alleviate the fronthaul infrastructure problem, we consider splitting the cell stack at the MAC layer.

---

[2] Connection to macro sites typically does not have dark fibre, but uses fibre that is shared (WDM) with other traffic.

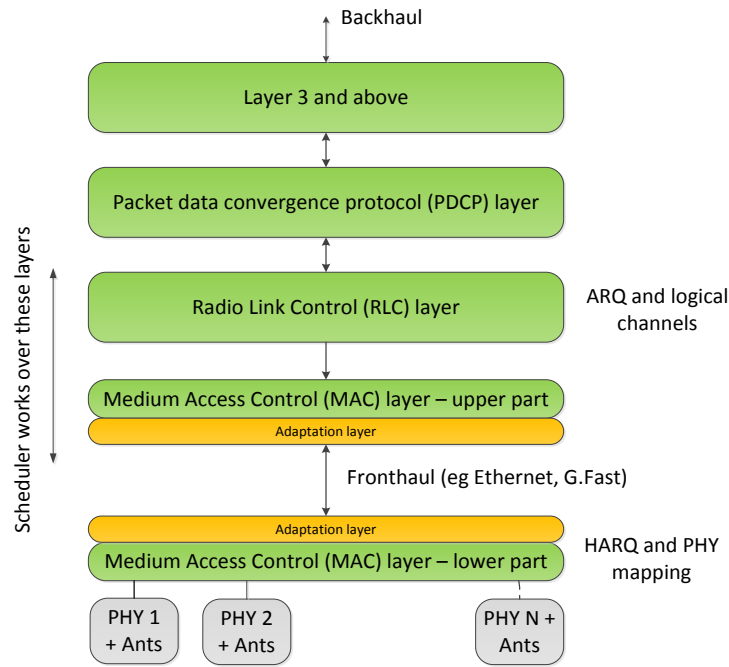

*Figure 10: Splitting at the MAC layer takes some pressure off the fronthaul but at the cost of some complexity*

In this case, the fronthaul traffic is approximately equal to the cell load, and tests done in the laboratory indicated an overhead of about 12% due to the adaptation over Ethernet. There is an increase in complexity with this method because of the need for adaptation layers interfacing Ethernet. It will also be necessary to study the effect of errors that occur on the fronthaul link. It is not clear yet in the project which MAC functions will be in the upper and lower parts, but if we assume that the scheduling takes place in the upper part. Then, centralisation of scheduling should enable more concurrent UEs to use the network, which is particularly beneficial for machine-type communications (e.g. with massive IoT).

The final place that we consider splitting for virtualisation is below the PDCP layer, as shown in Figure 11.

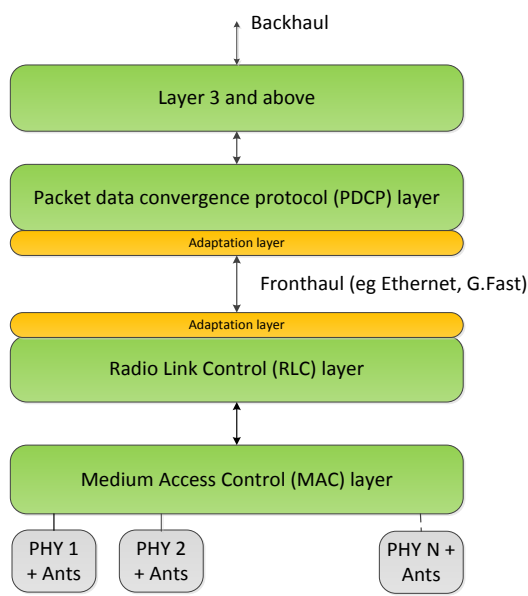

*Figure 11: Splitting at the PDCP layer is least problematic for the protocol stack but returns the least advantage*

 

Splitting here does not disturb at all the ARQ/HARQ mechanisms and it adds roughly the same overhead as splitting at the MAC layer. However, the benefit of virtualising is diminished because of the large amount of processing that is still being carried out in the remote part. The pros and cons of the split points are summarised in Table 2.

| Split point | Pros | Cons |
|---|---|---|
| PHY | Minimum remote equipment and maximum flexibility in VM parts | Large overhead, need for high bit-rate, low latency and time phase synch. Needs low latency links at a speed proportional to bandwidth and number of antennas. (e.g. needs dark fibre) |
| MAC | Synergy through centralising some scheduling. Small overhead, low sensitivity to latency | Need to spoof UL ACKs if latency is >8ms which may limit UL speeds. No correction of errors on UL, so re-transmissions need re-scheduling. Effects of errors on the fronthaul are unclear |
| PDCP | Small overhead, low sensitivity to latency | Low benefit since most of the processing is at the remote end. Effects of errors on the fronthaul are unclear |

*Table 2: Summary of cell split points and their pros and cons [21]*

**Summary**

Based on the analysis so far, the best place to split the cell functionality is at the MAC layer, since this provides sufficient benefits through virtualisation and yet is not too demanding on the fronthaul technology (capacity).This will provide synergy in SPEED-5G framework through centralising some MAC functionalities with reduce overhead and low latency. The disadvantages, which require further analysis, are reduction in uplink resilience and impact and propagation of errors that occur on the fronthaul. It may be possible to mitigate the effect of such fronthaul errors by incorporating error correction scheme into the adaptation layers.

## 4.4.2    Impact of fronthaul technology

In this subsection, we give a summary of copper technologies that are candidates for fronthaul especially from inside of buildings.  Most broadband connections in Europe have fibre to the cabinet (FTTC) with the tail to the premises using copper. In time this will migrate to fibre to the premises (FTTP).

The technology that runs over the tail from the cabinet to the premises is typically VDSL (ITU G993.1) or VDLS2 (ITU G993.2) and in the near future will be G.FAST (ITU G9701). Copper links are susceptible to impulse noise from electrical equipment, and this will introduce errors on the fronthaul, which VDSL does not attempt to correct, but G.FAST has error control options to mitigate these errors. These options in G.FAST can be disabled to obtain lowest latency; enabling error correction will add latency jitter to the G.FAST links. The decision to implement error correction on the fronthaul will depend on the impact of errors.

VDSL offers data rates of 55Mbit/s (DS) and 15Mbit/s (US), to 1000 metres copper length, and falls off with increasing distance. VDLS2 offers 100Mbit/s (DS) and 50Mbit/s (US) over 300m. With both VDSL and VDSL2, downlink and uplink occupy different parts of the spectrum and the communication is duplex. The latency over DSL is typically 10 – 20ms round trip.

G.FAST offers 300Mbit/s total DS and US (TDD format) over 300m distance and the split can be varied, not among individual houses but across a group that uses the same distribution card in the cabinet. Unlike VDSL, if the distance is reduced below 300m the rate climbs, reaching the limit of 1Gbits/s total DS and US at a distance of 100m. The latency of G.FAST is around 2ms each way minimum, when the error-correcting options are disabled but there will be latency jitter when errors are corrected through the HARQ mechanism.

More detail on Backhaul virtualization and its characterisation and functional split mapping can be found in Appendix A.

## 4.5 Virtual mobile small cells for ubiquitous high speed data services on demand

The evolution towards 5G is considered to be the convergence of internet services with existing mobile networking standards leading to the commonly used term "mobile internet" over heterogeneous networks (HetNets), with very high connectivity speeds. In addition, green communications seem to play a pivotal role in this evolutionary path with key mobile stakeholders driving momentum towards a greener society through cost-effective design approaches. In fact, it is becoming increasingly clear from new emerging services and technological trends that energy and cost per bit reduction, service ubiquity and high speed connectivity are becoming desirable traits for next generation networks.

Providing a step towards this vision, small cells are envisaged as the vehicle for ubiquitous 5G services providing cost-effective high speed communications. Pivotal to the 4G revolution is the well-known femtocell which is currently the market solution for providing energy-efficient high speed internet access for indoor scenarios. Complementary to femtocell technology, the LTE standard delivers the outdoor version in the form of picocell deployment suited for wide area coverage. However, the latter requires radio networking infrastructure and careful planning representing a significant cost for operators. Indoor femtocell technology is here to stay with a desirable energy rating making it a winning candidate for a basic building block on which to evolve mobile networks of the future.

Therefore, the question that arises is intriguing: what if we were to break the current mould of typical femto applications and extend femto accessibility to the outdoor world? Then perhaps we would stumble upon the next generation of femtocell technology for 5G networks. This is the idea hiding behind two contemporary solutions: fixed outdoor devices (metrocells) that provide femto-like services and tethering through mobile devices. Both of them are limited in speed, interoperability and coverage. To fully address this question, our work extends the notion of femto applications towards outdoor by employing virtual small cells. These small cells are set up on demand and constitute a "wireless network of cooperative small cells" that has a plethora of high speed backhaul connections to the mobile network. Moreover, network coding that has had strong application in augmenting network resiliency is used here as the overlay network to provide robust and cost-effective communications for supporting 5G services.

Virtual small cells not only need to be ubiquitous and cost-effective but have to deliver future emerging services in a secure fashion in an era where applications will handle "extremely confidential data" and money transactions. Therefore, 5G networks must deliver a framework with a palette of security tools as enablers for a cross-system end-to-end secure link that is fast and lightweight in nature.

Virtual small cells also open up the possibility for network operators to invest in network-sharing scenarios whereby operators can accommodate the foreseen increase in traffic whilst reducing their investment in new infrastructure and, beyond that, significantly reduce their energy bill. These enhancements are currently addressed by 3GPP, however, this raises new research challenges when applied to the virtual small environment. Following is the summary of our contributions.

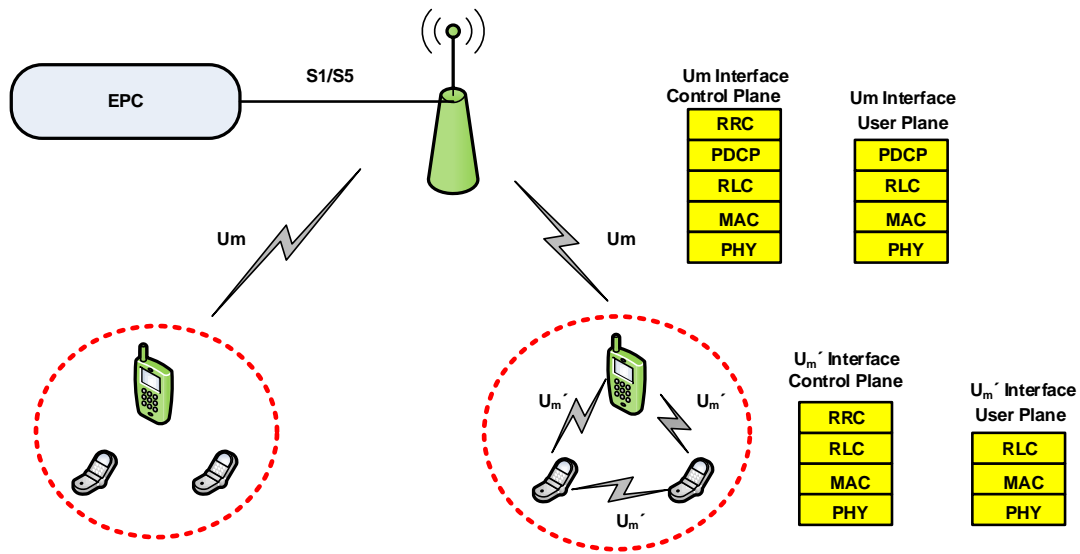

*Figure 12: Architecture for proposed on-demand small cells*

- Mobile small cells can be created according to traffic demand in hotspot area under the control of macro cell or standalone such as in disaster situation.
- These mobile small cells will play a very important role in the future network to meet new requirements in traffic volume, frequency efficiency, energy, cost, and so on.
- We will enhance the legacy 3GPP architecture for mobile small cells as shown in Figure 12.
- We will introduce new interface between eNB and mobile cell (when create), called Um. Interference protocol is shown in Figure 12, which is adopted from 3GPP.
- After a small cell is created, it acts as a virtual eNB for other users and helps to offload the traffic from macro eNB.
- Each small cell is also capable to schedule the users which are connected to them.

We use simplifying version of RRC in Um´ in control (C) plane and for user (U) plane we remove the PDCP layer to save terminal power.

## 4.6 Impact of the degrees of virtualisation on the resource management and MAC

Wireless network virtualisation inevitably places constraints on the radio resource management and MAC scheduler [4]. Resource allocation is a significant challenge of wireless network virtualisation. Resource allocation schemes need to decide how to map a virtual wireless network to physical networks resources (e.g. which nodes, links and resources should be picked and what should be optimised [5]). Resource allocation in a network virtualisation environment refers to a static or dynamic mapping of virtual nodes and links to physical nodes and paths, respectively. With constraints on resources or requirements, the resource allocation problem can be an optimisation problem with very high complexity. Unlike wired networks, resource allocation becomes much more complicated in wireless network virtualisation due to the variability of radio channels, user mobility, frequency reuse, power control, interference, coverage, roaming etc. [6]. Also, since the properties of uplink and downlink may not be the same in the wireless environment and the traffic is asymmetric in both directions, resource allocation should be considered for both uplink and downlink cases.

Resource scheduling is also important from the system performance perspective. As the range of services from SPs (Service Providers) can be very wide from best-effort to delay-sensitive, the QoS of these services must be dynamically mapped to physical wireless links. In order to be able to run on all elements of both virtual and physical elements, an efficient scheduling algorithm is needed to be implemented for both InPs and MVNOs.

The cooperation for radio resource management becomes important in radio virtualisation to achieve an efficient mapping of different wireless virtual networks in to a physical one. A MVNO can be simply assigned with a fixed set of Physical Resource Blocks (PRBs) such that each operator's traffic is scheduled only within its dedicated PRBs. An example of this static scheme with two operators is shown in Figure 13(a). PRBs are grouped with two fixed sets and each operator scheduler access only one of groups. However, due to restriction of frequency diversity, this solution provides poor overall spectral efficiency and unnecessarily limits the peak data rate available to users of one operator when there is low traffic in the cell of the other operator [5]. Considering these limitations, the MAC scheduler can dynamically allocate the required resource allocation for operators while monitoring the operator users' channel conditions and the amount of resource assigned to each operator. In this way, all operators have access to the whole system bandwidth as shown in Figure 13(b).

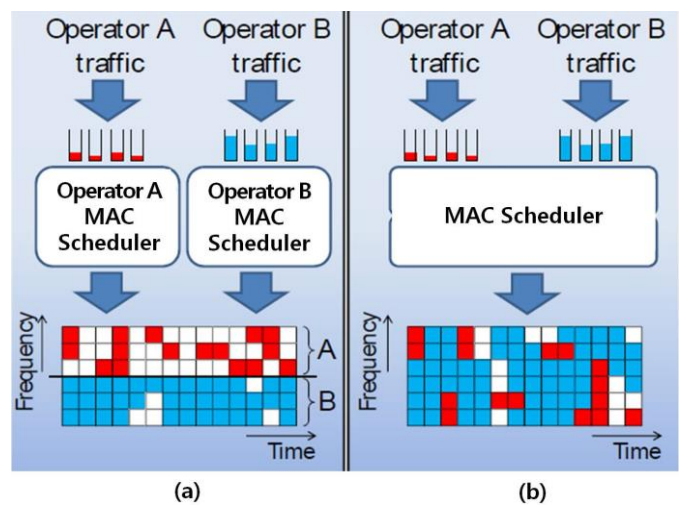

*Figure 13: An example of Radio Resource Scheduling in a virtualisation environment*

Additionally, depending on a virtual operator's request, a certain number of PRBs can be decided and allocated. While the traffic load is fluctuating over time, the amount of the required PRBs can vary depending on the traffic load of each operator. When PRBs are allocated to different virtual operators in a dynamic manner, at equal time intervals, considering traffic loads, each operator will get its required share of the PRBs and less waste of resources will occur. For efficient spectrum utilisation, investigation of an enhanced load estimation mechanism is required to determine each virtual operator's required bandwidth (i.e. PRBs) more accurately [7]. Since resource may be shared among multiple InPs, an efficient coordination mechanism should be designed appropriately.

   

Another issue in resource allocation is admission control. The objective of admission control is to maximise the utilisation while guaranteeing the QoS of existing users by controlling the admission of new incoming users. With wireless network virtualisation, in admission control for SPs, MVNOs need to conduct accurate traffic estimation and ensure that the virtual resources allocated to SPs do not exceed the capacity of underlying physical networks. This could be complicated in wireless environment because the number of end users and their traffic change dynamically in a certain geographic area, which causes unpredictable aggregated throughput in this area.

In resource allocation, the time granularity (i.e. how often should resource discovery and allocation be performed) can affect the system performance [4]. If the time interval is too small, the cost of overload and signalling may increase significantly. However, long time interval would lead degradation to static architecture of traditional networks.

Mobility management is an important issue in wireless networks that ensures successful delivery of new communications to users and maintains ongoing communication with minimal disruptions, while users move [8]. There are two components in mobility management: location management and handover management. Location management enables the network to deliver communications to users by tracking their locations. Handover management provides service continuity by keeping a user connected when its point of connection to the network moves from one base station to another. With wireless network virtualisation, tracking a user's location could be challenging, since a user's location update may need to be performed with different MVNOs or InPs (Infrastructure Providers). A centralised location management can solve the problem. However, latency will be introduced in centralised management, thus some distributed mechanisms could be investigated. In addition, since a user with ongoing communications may switch among multiple MVNOs or INPs, the handover management problem becomes more complicated than that in traditional wireless networks [6].

## 4.7    Aspects of virtualisation – benefits and performance metrics

In this section, the main metrics that can be used to evaluate the performance of a virtualised wireless network will be described. It should be noted that the specific metrics can be employed to compare different virtualisation architectures as well as different resource allocation mechanisms. The entire set of performance metrics can be separated into two metric sets: the first one refers to the performance metrics that are also used in traditional wireless networks whereas the second set consists of virtualisation-specific metrics [6]. To be able to investigate the performance of a virtualised network/architecture both types of metrics can be employed, however the use of virtualisation-specific metrics or their joint consideration in case of a resource management problem gives further insight to the virtualisation process.

### 4.7.1    Performance metrics of traditional wireless networks

*Costs (CapEx, OpEx)*: This metric refers to the total costs regarding the infrastructure constructing and the operation/maintenance of the wireless network. More specifically, these costs can be separated in three main categories: i) the license costs for the use of the spectral resources (that may be considered as a CapEx cost component), ii) the CapEx costs, referring to the costs of the base station (BS) equipment, radio network controller (RNC) equipment, core network (CN) equipment and iii) the OpEx costs, referring to energy charge, site/backhaul lease, operation and maintenance costs. Network virtualisation will result in additional costs regarding both equipment and operation/maintenance. However, the direct comparison of total costs in traditional wireless networks and virtualised wireless networks may not be representative of the virtualisation benefits, as it gives no insight for the additional flexibility and resource utilisation that is provided in the virtualised networks. To be able to take into account these benefits, we should also consider the generated revenues in each case and to compare the corresponding relation of revenues/costs. Depending on the virtualisation architecture, the revenues may represent the income from providing services to the users or from leasing infrastructure to the service provider. Towards this direction, there are various utility functions that can be defined in order to investigate the network performance either using the profit function [9][10] or the revenue to cost ratio (RCR) function. The first metric is defined as the difference between the revenue and the cost whereas the second one is defined as the ratio of the revenue to the respective cost. The relation between these metrics and the virtualisation benefits is monotonically increasing, meaning that the higher the values of these metrics, the higher the motivation for the providers to deploy virtualisation mechanisms.

***Deployment efficiency***: This metric is defined as the ratio between the achieved system throughput and deployment costs (CapEx+OpEx) and it constitutes a significant network performance indicator. In case of virtualised wireless networks, the specific metric can be used as an additional metric to evaluate the benefits from the virtualisation.

***Energy efficiency***: This metric can be defined either as the throughput per energy ratio or the throughput per power consumption ratio. In both cases, it is expected that virtualised wireless networks will offer high degrees of energy efficiency compared to the traditional wireless networks by increasing the system throughput and also by reducing the energy consumption of the system (using advanced power saving solutions/sleep configurations).

***Delay & jitter***: The metric of delay refers to the requested time for a packet to be transferred from one node to another node in the network, whereas jitter is employed to measure the packet inter arrival times. In case of virtualisation networks, these metrics may refer to the corresponding times among the virtual nodes of the network. It should be noted that both metrics can be used for the evaluation of the resource management mechanisms and the virtualisation process/architecture.

***Capacity/Throughput***: This capacity constitutes an important metric of the traditional wireless systems and it is employed in order to investigate the rate that can be achieved using the specific network architecture and a specific resource management scheme. In a similar way, in case of virtualised wireless networks, this metric can be employed in order to evaluate/compare different architectures and different allocation schemes.

***Effective capacity***: This metric can be defined as the maximum rate that can be sustained in order to guarantee specific QoS requirements to the users [11]. It constitutes a useful tool that is employed in traditional wireless networks as it can integrate the rate performance of a wireless link and the key metric of delay QoS requirements. Similarly, in the case of virtualised networks, this metric is particularly convenient for the analysis of the statistical QoS performance to investigate different architectures and resource allocation schemes (mainly power allocation mechanisms).

### 4.7.2 Virtualisation-specific metrics

***Virtualisation-specific throughput***: This metric refers to the average throughput among the virtual entities and it can be employed in order to measure the connection performance either between virtual nodes or between service providers/network operators to end users and to evaluate various resource management schemes in virtualised wireless networks.

***Network Utilisation***: Utilisation is defined as the ratio of the used substrate resources to the total amount of available resources. The higher is this metric, the higher the accumulated benefits from the virtualisation process. Utilisation can also be employed in order to evaluate the performance of different resource management mechanisms and virtualisation architectures.

***Path length between virtual entities***: This metric is defined as the number of links between two sub-network nodes that are mapped with one direct virtual link connecting them. The main reason that this metric is investigated is due to its impact to the delay and jitter metrics. Specifically, longer path length results in higher delays/jitter as there are more nodes that have to forward the packet. This metric is appropriate for evaluating and comparing the different virtualisation architectures as well as the different resource allocation schemes.

Finally, another virtualisation-specific metric that can be used is the isolation level that refers to the lowest virtualised physical resource level. However, this metric is not always indicative of the network performance. More specifically, given that isolation among different virtualised wireless networks can be done at different levels (flow-level, physical resource level, infrastructure-level etc), there is a trade-off among implementation complexity and performance. In general, isolation at a higher level may be simpler for implementation at the cost however of an inefficient allocation or non-strict isolation. On the other hand, isolation at a lower level may achieve much better resource utilisation and system throughput at the cost of high computational complexity. Hence, the isolation level has to be chosen depending on the performance requirements of the virtualised network.

## 4.8    Network virtualisation standardisation

Network Virtualisation (NV) has found success in wired networks for decades to abstract physical network resources into logical networks, with the latter providing end-to-end services. By abstracting the logical network behaviour from the physical network resources, each of these could be managed independently leading to economies of scale for Service providers (SPs) and also breaking the monopoly of big companies in this industry, which in-turn creates a healthy competitive environment. 3GPP TR 32.842 [12] elaborates on the advantages NV has over the traditional service provisioning through the physical and dedicated network resources, which include cost reduction, reduce time to market, reduced operation cost and on-demand flexible service provision.

It is imperative that standardisation bodies are involved in NV, shaping the direction of the requirements, innovations and services that NV can support to facilitate cooperation and industry-wide adoption. ETSI has embarked on such an activity, by setting up the NFV Industry Specification Group (NFV ISG) [13]. The objective of this Specification Group is not to standardise NV per se, rather to define use cases, requirements and architectural framework, which become the base on which businesses and industry can influence NV. In addition, they provide a platform to demonstrate any proof of concept (PoC).

Figure 14 illustrates an NV architecture framework proposed by ETSI for industry-wide adoption.
- **Management and Network Operations** (MANO):
    - **VNF Manager**: This entity manages VNF (Virtualised Network Functions**)** lifecycle that include its instantiation, configuration and other operational functionalities, as well as coordination among VNFs.
    - **Virtualised Infrastructure Manager**: Manages the actual hardware and software resources, which include networking, storage and computing resources.
    - **Orchestrator**: This is the heart of the MANO, managing and coordinating the functionalities of the VNF Managers and the Virtualised Infrastructure Manager. It also interfaces to the OSS/BSS, facilitating seamless integration into traditional non-virtualised network management systems.
- **Network Function Virtualised Infrastructure** (NFVI): NFVI consists of the actual physical hardware resources, which include the commercial-off-the-shelf (COTS) computing, storage and network resources and the other part is the virtualised versions that enable independent VNFs to share the resources.
- **Virtualised Network Functions** (VNF): This is the virtualised version of a Network Function (NF), which traditionally would have been tied to a specific hardware.

Cellular Network Virtualisation has received huge attention lately, with Mobile Network Operators (MNOs) looking to leverage the advantages of network virtualisation. In Cellular communications, e.g. LTE, the different entities such as the core network (CN), radio access network (RAN), spectrum and even the devices (e.g. D2D) can all be virtualised albeit the challenges. This opens up the cellular market for other players to contribute.

3GPP, the sole standardisation body for cellular communications, has backed ETSI NFV ISG's NV architecture framework, use cases and requirements to come out with a reference model that can be adopted by MNOs [12]. This activity rather focused only on the CN. This leaves room for contributions in RAN, spectrum and devices virtualisation, all in the quest of meeting 5G requirements.

SPEED-5G's virtualisation techniques, with strong focus on RAN, will refine the direction the standardisation bodies take in RAN and spectrum (e.g. eDSA) virtualisation.

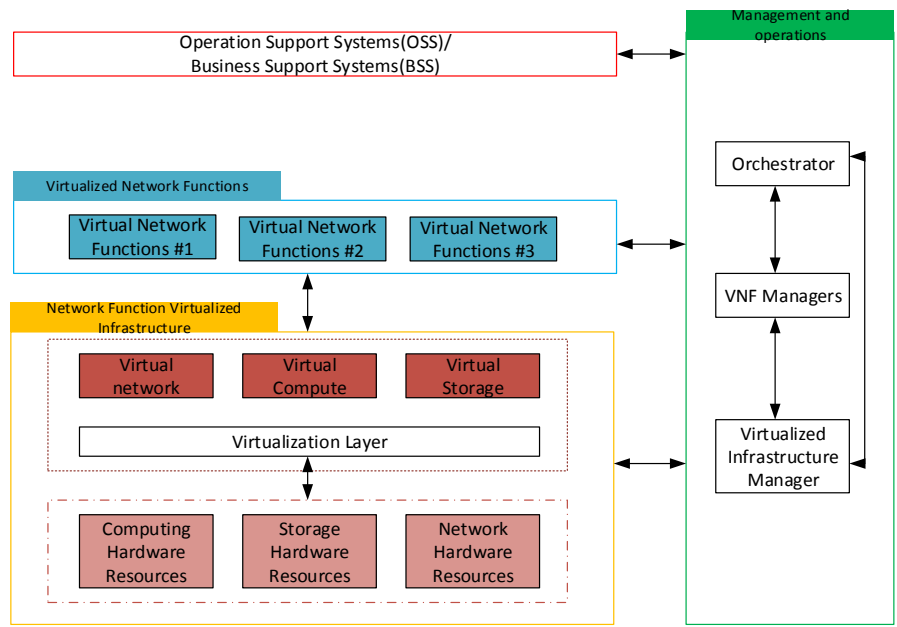

*Figure 14: NFV Architectural Framework [14]*

# 5 SPEED-5G Performance Metrics

This section provides an overview of the key performance indicators (KPIs) identified to assess the performance of the technical solutions designed within SPEED-5G. The 5G-PPP high level KPIs will be first presented to provide the appropriate framework for the planned contributions of SPEED-5G, and then KPIs of SPEED-5G KPIs will be elaborated for different use cases as described in Section 2.1. Due to the wide variation of the environmental conditions in the different SPEED5G use cases, there is also a corresponding spread of KPIs' values which has to be taken into account, i.e., a single KPI value will not usually fit all use cases considered in SPEED-5G.

## 5.1 Relation to 5G-PPP KPIs

On 2013, the European Commission and the 5G infrastructure PPP signed the PPP Contractual Arrangement [15], which, among other aspects, describes the high-level KPIs of the PPP for the period starting in 2014 and finalizing in 2020. In this context, as one of the funded projects within the 5G-PPP, SPEED-5G's planned contributions constitute a step forward towards achieving those KPIs, as shown in Table 3 .

| KPI | Relevance (High/Medium/Low) | Details on planned SPEED-5G's contribution towards achieving the KPI |
|---|---|---|
| **Performance KPIs** | | |
| Providing 1000 times higher wireless area capacity and more varied service capabilities compared to 2010 | High | SPEED-5G implements extended-DSA (eDSA) concept towards achieving higher wireless capacity based on the following three functions: 1) better resource reuse along small-cell based ultra-densification; 2) traffic smart offload with the use of heterogeneous technologies; and 3) efficient use of all and any available spectrum resources through the use of dynamic and smart spectrum access. |
| Facilitating very dense deployments of wireless communication links to connect over 7 trillion wireless devices serving over 7 billion people | High | SPEED-5G contribution in increasing the density of connected devices will be achieved via following approaches: 1) increasing the overall simultaneously operated bandwidth (via the use of licensed and unlicensed spectrum at the same time); 2) allowing more users per spatial area (e.g., by supporting dynamic channel selection to mitigate interference); 3) adding small cells that can offload traffic; and finally 4) optimizing the medium access control protocol so that users can access the medium more efficiently (i.e., based on consideration of current traffic loads on channels, traffic types to support, different regulation regimes and so on). |
| Creating a secure, reliable and dependable Internet with a "zero perceived" downtime for services provision | Low | In SPEED-5G, reliability of Internet connectivity will be improved by addressing service continuity. This will be achieved through new approach to rationalise the management coexistence and cooperation of wireless technologies to avoid the spectrum crunch, by breaking the technology silos (smart offload, and dynamic spectrum access) prioritizing traffic according to available resource and technology capabilities. |
| **Societal KPIs** | | |
| Stimulation of new economically-viable services of | High | SPEED-5G addresses this aspect via stimulation of M2M applications, more precisely, sensor based M2M |

| high societal value like U-HDTV and M2M applications | | applications like gas metre. In SPEED-5G, in order to implement various M2M applications, new MAC algorithms for IoT traffic will be designed with consideration of the use of FBMC and the unlicensed spectrum. |
| --- | --- | --- |

*Table 3: 5GPPP KPI list*

In the following three subsections, the 5G-PPP KPIs described in [15] that are relevant to SPEED-5G project are presented. For each 5G-PPP KPI, a qualitative assessment of the relevant of that KPI in relation to the contributions provided in SPEED-5G is also provided together with details on the planned SPEED-5G contribution towards the achievement of the KPI.

## 5.2    SPEED-5G high level KPIs

In order to be able to provide quantifiable contributions from the SPEED-5G project towards the achievement of the 5G-PPP KPIs described in the previous section, one of the activity within this deliverable is to define and quantise a set of high level KPIs, that, in the following, will be referred to as SPEED-5G KPIs.

Given the wide scope of the research topics addressed in SPEED-5G, these derived high level KPIs will then be particularised in different tasks for each one of the three technology areas addressed in SPEED-5G (each one corresponding to the deliverable):

- RM framework and modelling

- MAC approaches with FBMC

- Testbed deployment and trials

These particularisations will yield to specific design principles, requirements, and guidelines for the development of the corresponding enabling technologies in each one of those areas. These particularisations are especially important in this initial phase of the 5G PPP because they will set the starting point upon which all the technology developments will capitalise. Thus, it was of paramount importance for the proper development of the SPEED-5G project activities to choose a set of high-level KPIs that adequately capture the requirements expected from 5G communication platforms for the subset of use cases described in the previous section 2.1 and that, also very importantly, allow proving the level of fulfilment towards achieving the 5G-PPP KPIs described in Section 5.1.

Discussions started on which should be the list of high level KPIs, both internally, at consortium level, but also externally with other 5G-PPP projects. The outcome of these discussions, the so-called "Consolidated KPIs" is presented in following section together with an explanation on the meaning of each KPI and how SPEED-5G intends to tackle it.

### 5.2.1    Data rate

**User-experienced data rate:**
- *Definition:* It is defined as "Throughput, measured in bit/s at the application layer"
- *Evaluation method*: Simulation

**Average Service Throughput per Cell:**
- *Definition*: The average service throughput per cell is defined as the sum of the total amount of bits successfully received by all active users in the system, divided by the product of the number of cells simulated and the simulation duration

- *Evaluation method*: Simulation

**User Average Peak Bit rate:**
- *Definition*: The highest theoretical data rate which is the received data bits assuming error-free conditions assignable to a single mobile station, when all available radio resources for the corresponding link direction are utilised (i.e., excluding radio resources that are used for physical layer synchronisation, reference signals or pilots, guard bands and guard times)
- *Evaluation method*: Simulation/Analytical

**Per-user Service Data Throughput:**
- *Definition*: The user's service data throughput is defined as the ratio of the number of information bits successfully received by the user and the total simulation run time.
- *Evaluation method*: Simulation

**Cell-edge user throughput:**
- *Definition*: The cell edge user throughput is defined as the fifth percentile point of the CDF of user's average packet call throughput
- *Evaluation method*: Simulation

**SPEED-5G approach**
This KPI relates to the achieved end-user data rates (both UL and DL) in different forms: User experience data rates, average service throughput per cell, user average peak bit rate, per user service data throughput and cell edge user throughput.

For example, SPEED-5G will contribute to increase the typical data rate by relying on a combination of densification of small cells, efficient use of spectrum and exploiting multi-RAT resource management. Resource management is being applied in an effective way only to single technologies (e.g. LTE) but not to multi-RAT environments. At the core of the project is the definition of a new MAC layer that facilitates multi-RAT access, and allows prioritising and allocating traffic across heterogeneous access technologies.

## 5.2.2 Latency

**E2E Latency:**
- *Definition*: total delay between server and client including video decoding
- *Evaluation method*: Testbed

**Link Latency:**
- *Definition*: Duration of a packet from user terminal to the Layer 2 / Layer 3 interface of the 5G system destination node OR The amount of time a packet stays in the eNB queue until correctly received at the UE
- *Evaluation method*: Simulation

**SPEED-5G approach**
This KPI relates to the network latency (round trip time) and to the link latency, which is measured as the time between a packet being available at the transmitter and the availability of this packet at the receiver (which takes into account, e.g., constraints and delays imposed by the HW).

SPEED-5G will provide solutions for 5G communication platforms so that the latency can be reduced by proposing proposed architecture follows the most recent software trends, where traditional networks are replaced by logical networks deployed over slices. The novel slice concept is based on deploying logical elements that can easily be defined, modified and started up thanks to the principles of Software-Defined Networks (SDN) which allows configuring and instantiating the SW by decoupling control from data plane; Network Function Virtualisation (NFV) that supports virtual HW abstraction mechanisms and; Software-Defined Radio (SDR) that supports the stack virtualisation procedures. The proposed architecture reduces the control signalling, improves throughput, reduces latency for supporting real-time services, enhances spectrum access without interfering with other systems, and finally provides flexibility, scalability and an easy interconnectivity with other networks.

### 5.2.3    Energy

**Energy efficiency:**
- *Definition*: The number of bits that can be transmitted per Joule of energy, computed over the whole network
- *Evaluation method*: Simulation

**Energy consumption:**
- *Definition:* The amount of energy (W) consumed in a $Km^2$
- *Evaluation method*: Simulation

**Node lifetime:**
- *Definition:* Expected battery lifetime
- *Evaluation method*: Simulation

**Idle lifetime:**
- *Definition:* Duration an IoT node is in idle time
- *Evaluation method*: Simulation

**SPEED-5G approach**
SPEED-5G will provide solutions for 5G communication platforms so that the energy efficiency/consumption can be reduced. One of the major designs principles is to reduce the consumption of the radiated energy and the operating energy is to adjust the capacity of the network to the demand. Also, mechanisms for better scheduling or transmission power control schemes will help in a power reduction combined with algorithms that enable the base station to get into the sleep mode more effectively, thus saving energy. Also, SPEED-5G is proposing the novel idea of D2D using LAA/LTE-U, novel channel selection scheme [18]-[19] and small-cell-based load balancing via LAA/LTE-U type mechanism.

### 5.2.4    Spectrum

**Spectrum efficiency:**
- *Definition*: The peak data rate normalised by bandwidth
- *Evaluation method*: Simulation

**Bandwidth flexibility:**
- *Definition*: The ability of the access technology to operate with different bandwidth allocations
- *Evaluation method*: Simulation

**SPEED-5G approach**

The SPEED-5G project proposes a novel architecture focussed on small cells and their particular characteristics and challenges. The challenges are mainly related to the efficient use of spectrum and its management key features. These functionalities need to be addressed by design.

In order to satisfy the capacity requirements of future wireless services and maximise the macro cell offloading, solutions able to provide more efficient spectrum usage are required. On the one hand, it is necessary to take full advantage of available spectrum by dynamically access, licensed, licensed shared, and license-exempt bands. On the other hand, enhanced interference coordination mechanisms have to be designed to enable fair coexistence on frequency resources and maintaining high QoE. To achieve these goals and reduce the complexity of network planning and deployment phases, the envisioned 5G architecture should allow for easy and autonomous management, configuration, and optimisation of small cells. Finally, a lean and scalable design characterised by limited interfaces is required to easily manage a high number of small cells, avoid signaling congestion, and improve the system efficiency.

### 5.2.5    Reliability

**Reliability Rate:**
- *Definition*: The amount of sent packets successfully delivered to the destination within the time constraint required by the targeted service, divided by the total number of sent packets
- *Evaluation method*: Simulation

**Packet Loss Ratio:**
- *Definition*: The percentage of packets lost with respect to packets sent.
- *Evaluation method*: Simulation

**Downlink availability:**
- *Definition*: DL packet transmission opportunity rate
- *Evaluation method*: Simulation

**Fairness:**
- *Definition*: the Gini Index value lies between 0 and 1 if we go toward line of perfect fairness (closer to 0) our fairness increase.  0 means perfect Fairness if we go away from line of perfect fairness (closer to 1) our fairness decreases.
- *Evaluation method*: Simulation

**Mobility:**
- *Definition*: Movement of users inside the whole network area.
- *Evaluation method*: Simulation

**SPEED-5G approach**
The SPEED-5G project proposes a novel algorithm/protocol by considering the previously mentioned use cases, where these metrics can be useful in order to calculate reliability, packet loss ratio, fairness, and mobility. The required data that is needed for the users in the various use cases has to be received in the required time and not be dependent on the technology used. In particular reliability is becoming more critical as we want mobile communications for control and safety of machines. For example, in Machine-to-Machine (M2M) communications reliability is a must because automated machines can be controlled from great distances to do different tasks in workspaces with people around them. The term "reliable" is a synonym for "assured" meaning that the network must deliver a message/ packets with a minimum amount of packet losses which implies little delay for the package to be received after the process of decoding.

### 5.2.6    Connection

**Connection density:**
- *Definition*: Number of simultaneous devices that are exchanging data with the network per square kilometre
- *Evaluation method*: Simulation

**Duplexing flexibility:**
- *Definition: The ability of the access technology to adapt its allocation of resources flexibly for uplink and downlink for both paired and unpaired frequency bands.*
- *Evaluation method*: Simulation

**SPEED-5G approach**
The SPEED-5G project proposes a novel algorithm/protocol by considering previous mentioned use cases, where these metrics can be useful in order to connection density and flexible duplexing mode. The market shows that while smart phones are expected to remain as the main personal devices and stay at almost the same numbers, but the total of other kinds of devices, including wearable devices and MTC devices, will continuously increase. So to meet the service and market demand towards year 2020 and beyond SPEED-5G will take into account these provisioned numbers and try to steer the technology into that direction, providing as the KPIs explain thousands of users per square kilometre with the connection needed.

## 5.3    QoE

**PSNR (Peak SNR):**
- *Definition*: For the three components[3] *Y*, *Cb* and *Cr*, the PSNR separately indicates the logarithmic ratio of the maximum possible total deviation of all pixels from the nominal value of the reference (MAX)
- *Evaluation method*: Testbed

**Picture freeze:**
- *Definition: repetition of one picture over 4 or more picture periods*
- *Evaluation method*: Testbed

**MOS-V (Mean opinion score video):**
- *Definition: based on ITU-T BT.500 quality in 5 steps (Bad, Poor, Fair, Good, Excellent)*
- *Evaluation method*: Testbed

**SPEED-5G approach**
By implementing eDSA concept, SPEED-5G plans optimised resource utilisation across heterogeneous technologies and spectrum of multiple regimes. However, heterogeneity of technologies and quality of spectrum will impact quality of experience delivery. SPEED-5G will analyse the impact of spectrum allocation and interference limitations in ultra-dense system on quality of experience. Since video represents 70% of mobile traffic, video traffic will be used as a reference to evaluate the proposed algorithms based on identified QoE parameters. QoE evaluation will be conducted based on a test platform for the measurement of video quality.

---

[3] For the three components of a video picture (luminance Y and the color difference values Cb and Cr), the Peak Signal to Noise Ratio (PSNR)

### 5.3.1 Backhaul

**PtMP backhaul link latency**
- *Definition*: The delay between the central node and a terminal node that a frame faces in a point-to-multipoint backhaul topology
- *Requirement:* $\simeq$ 1ms
- *Evaluation method:* Testbed

**PtMP backhaul link data rate**
- *Definition*: The aggregate data rate between the central node and all terminal nodes in a point-to-multipoint backhaul topology
- *Requirement:* $\simeq$ 2.5 Gbps (DS), $\simeq$ 2.0 Gbps (US)
- *Evaluation method:* TestbedF

**PtMP backhaul system availability**
- *Definition:* The percentage of time the PtMP system is operational
- *Requirement:* 99.999% (5-nines)
- *Evaluation method:* Analysis

**PtMP backhaul resource balancing**
- *Definition:* In a point-to-multipoint system using two central nodes in 1:1 mode, this is the capability to assign a terminal to one of the two central nodes based on their workload status
- *Requirement:* the objective is to achieve a fair balancing between the two hubs
- *Evaluation method:* Testbed

**SyncE: EEC output frequency accuracy for PtMP backhaul link**
- *Definition:* The frequency accuracy of the Synchronous Ethernet Equipment Clock (EEC), according to ITU-T G.8262/Y.1362, measured at the ports of a PtMP backhaul
- *Requirement*: ≤ 4.6 ppm
- *Evaluation method* : Testbed

**IEEE 1588v2: CF accuracy for PtMP backhaul TC**
- *Definition:* The accuracy of the Correction Field (CF) added in IEEE 1588v2 frames by a Transparent Clock (TC), according to ITU-T G.8273.3
- *Requirement*: ≤ 80 ns
- *Evaluation method* : Testbed

**SPEED-5G approach**
Within the SPEED-5G project, we focus on strengthening a weak link in an envisioned 5G network: the PtMP backhaul technology, whose cost advantage over fibre and PtP wireless alternatives as well as its installation ease, will continue to render it important to the operators. Specifically, our effort aims to improve the performance of a PtMP backhaul system towards 5G KPIs with focus on latency, data rate, resource balancing, high-availability and synchronisation accuracy. The target is to bring latency down to approximately 1ms and at the same time increase the rate to approximately 2.5Gbps (DS) and 2.0Gbps (US). The backhaul system high-availability and resource balancing will be addressed through redundancy and a novel protocol that will control the resources. Finally, the system will be enhanced with SyncE and IEEE1588v2 network synchronisation standards to meet strict accuracy requirements for supporting synchronous-sensitive end-to-end applications. We elaborate on these subjects in Appendix A (KPI Backhaul).

## 5.4    KPI Targets for use cases

Out of a wide range of available metrics there are some Key Performance Indicators (KPIs) that SPEED-5G is interested in monitoring. These can be broken down into a variety of categories and especially in some use cases that the project will be involved with. The SPEED-5G project has evolved over time to create a set of common values (aka Best Practices) that should be monitored. Below is an overview of some of these values broken into common categories. For example, the table below includes target KPIs related to latency, energy consumption, reliability, etc. for the different use cases and of course their related requirements that take into account all the bits and pieces that these cases demand in order to provide the best possible service and the user needs. Mainly these target KPIs are based on NGMN and other industry standards explained in a way that the SPEED-5G project considers to be comprehensible in order to meet the market target of the year 2020.

| Scenario | Use cases | KPI | Requirements | Notes |
|---|---|---|---|---|
| **Broadband Wireless** | UC1, UC2, UC3 | Connection Density | 200-2500 users/km$^2$ | In dense areas |
| | | User-experienced Data Rate | DL: up to 300Mbps UL: 50 Mbps | 5-8 Mbit/s for Full-HD 20-35 Mbit/s for 4k video 50-90 Mbit/s for 8k video |
| | | Cell-edge user throughput | DL: 4 Mbps | |
| | | U-plane Latency | <5s | For live TV viewing: < 5s For gaming: 50ms For live Outside Broadcast production: 1s |
| | | Energy efficiency | 50 joule/Mbps | |
| | | Energy consumption | 3 kW/km$^2$ | 15 watts maximum for residential base station |
| | | Reliability Rate | min 95% | |
| | | Packet Loss Ratio | max 5% | 16Mbit files |
| | | Urban mobility | On demand | Broadcast Pedestrian |
| **Massive IoT** | UC1, UC2, UC3 | Connection Density | 200k devices/km$^2$ | In dense areas |
| | | Minimum guaranteed data rate | DL: 10 kbps - 1 Mbps UL: 10 kbps - 1 Mbps | |
| | | E2E Latency | < 1 s | |
| | | Energy consumption | Batteries to last tens of years | |
| | | Urban mobility | Zero and limited | Zero: pedestrian (0-3km/h) Limited: 95% at zero and 5% at 0-120km/h |
| **Ultra-reliable** | UC1, UC2, UC3 | Connection Density | 100 /m$^2$ | Hotspots in a hospital (but varies a lot) |
| | | User Experience Data Rate | DL: up to 300Mbps UL: up to 300Mbps | For multiple video streams (e.g. Remote monitoring & case) |
| | | E2E Latency | 100ms < but <1s | The generic case (much tighter for robotic operation) |
| | | Availability | 99.999% | 99.99999 in case of robotics |
| | | Mobility | Up to 250km/h | 250km/h for high speed vehicles (500km/h in case of intervention in a helicopter) |
| | | BLER | 10$^{-6}$ | After MAC processing for video |
| | | Connection Density | 200-2500 users/km$^2$ | In dense areas |
| | | User Experience Data Rate | DL: up to 300Mbps UL: 50 Mbps | 5-8 Mbit/s for Full-HD 20-35 Mbit/s for 4k video 50-90 Mbit/s for 8k video |
| | | Cell edge user throughput | DL: 4 Mbps | |

| High Speed | UC1, UC3 | U-plane Latency | <5s | For live TV viewing: < 5s<br>For gaming: 50ms<br>For live Outside Broadcast production: 1s |
|---|---|---|---|---|
| | | Energy efficiency | 50 joule/Mbps | |
| | | Energy consumption | 3 kW/km$^2$ | 15 watts maximum for residential basestation |
| | | Reliability Rate | min 95% | |
| | | Packet Loss Ratio | max 5% | 16Mbit files |
| | | Urban mobility | On demand | Broadcast<br>Pedestrian |

*Table 4: Summary of main requirements and KPIs for different SPEED-5G use cases*

# 6    Conclusions

This deliverable presented the three realistic and demonstrable use cases considered in SPEED-5G. The deliverable has also outlined the SPEED-5G architecture design principles which are based on flexibility, simplicity, on-demand resource allocation, auto-scaling and enhanced performance through the extensive utilisation of software-defined networking and network function virtualisation. The degrees of virtualisation are elaborated in terms of virtual small cell deployment options, virtualisation of fronthaul/backhaul, virtualisation impacts on RRM/MAC design, and virtualization-related KPIs. Finally, the relation between 5G-PPP KPIs and SPEED-5G KPIs is presented and various KPIs identified for different SPEED-5G use cases are described. Potential solutions developed during the project will be benchmarked against the identified KPIs in this deliverable. In the next step, details of each of the identified use cases, refinement of function processes in terms of RRM/MAC will be provided in D4.1 (WP4) and D5.1 (WP5) deliverables.

# Appendix A    KPI: Fairness using Gini-Coefficient

The Gini coefficient was developed by the Italian Statistician Corrado Gini (1912) [24] as a summary measure of income inequality in society. It is usually associated with the plot of wealth concentration introduced a few years earlier by Max Lorenz (1905). Since these measures were introduced, they have been applied to topics other than income and wealth, but mostly within Economics (Cowell, 1995, 2000; Jenkins, 1991; Sen, 1973).

G is a measure of inequality, defined as the mean of absolute differences between all pairs of individuals for some measure. The minimum value is 0 when all measurements are equal and the theoretical maximum is 1 for an infinitely large set of observations where all measurements but one has a value of 0, which is the ultimate inequality (Stuart and Ord, 1994). When G is based on the Lorenz curve of income distribution, it can be interpreted as the expected income gap between two individuals randomly selected from the population (Sen, 1973). The Lorenz curve is plotted as the cumulative proportion of the variable against the cumulative proportion of the sample (i.e. for a sample of 30 observations the cumulative proportion of the sample for the 15th observation is simply 15/30). To get the cumulative proportion of the variable, first sort the observations in ascending order and sum the observations, then each kth cumulative proportion is the sum of all $x_i/x_{sum}$ from i=1 to k.

The classical definition of G appears in the notation of the theory of relative mean difference as shown in equation:

$$G = \frac{\sum_{i=1}^{n}\sum_{j=1}^{n}\left|x_i - x_j\right|}{2n^2\bar{x}}$$

- where x is an observed value, n is the number of values observed and x bar is the mean value.

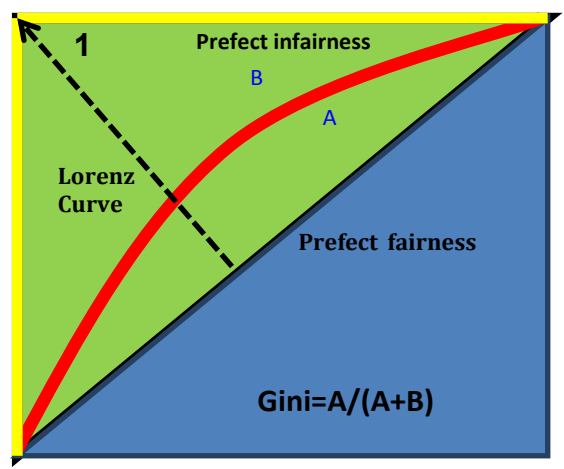

*Figure 15: Lorenz Curve*

Gini Index value lies between 0 and 1 if we go toward line of prefect  fairness    (closer to 0) our fairness increase.0 mean prefect Fairness if we go away from line of prefect fairness(closer to 1) our fairness decreases. Prefect in-fairness Lorenz Curve tells us how much data is deviated from the Line of Prefect fairness. The Gini coefficient is defined graphically as a ratio of two surfaces involving the summation of all vertical deviations between the Lorenz curve and the perfect Fairness line (A) divided by the difference between the perfect Fairness and perfect unfairness lines (A+B).
SNR Based Fairness using Gini Formula

$$G = \frac{\sum_{i=1}^{n} \sum_{j=1}^{n} \left| SNR_i - SNR_j \right|}{2n^2 \, \overline{SNR}}$$

*where*

*'SNR'  observed  SNR value.*

*'n'  is the number of  SNR oberved .*

$\overline{SNR}$  *is the mean value*

# Appendix B    KPI: Backhaul

Backhaul network is considered an integral part of the network architecture in 5G [6]. Its performance can affect the end-to-end performance of the 5G services and thus will be also taken into account in this study.

Figure 16 depicts the position of the backhaul network within 2G/3G/4G mobile networks. It is the part of the network that comprises the intermediate links between the Core Network and the small sub-networks at the "edge" of the entire hierarchical network (cells). Backhaul plays a vital role in mobile networks by acting as the link between Radio Access Network (RAN) equipment and the mobile backbone, by transferring voice and data from the access base stations to the core network.

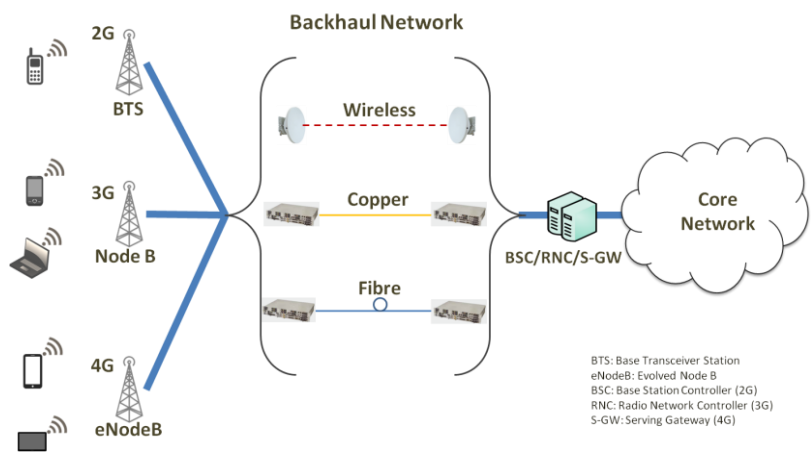

*Figure 16: Backhaul in 2G/3G/4G Networks*

There are a wide variety of technologies and solutions available as carriers for backhauling traffic. Mobile backhaul can be performed via fibre, copper or wireless links. Operators might adopt more than one technology deploying each where most appropriate. It's critical though for the backhaul network to meet specific cost, coverage and capacity objectives, without compromising service quality.

## Backhaul characterisation and functional split mapping

The EU project iJOIN has analysed [16] the characteristics of different BH technologies to understand their impact on the Radio Access Network performance. Table 5 describes the result of these studies where BH types are differentiated in terms of latency, throughput, topology, duplexing and multiplexing technologies.

| Number | BH technology | | Latency (per hop, RTT) | Throughput | Topology | Duplexing | Multiplexing Technology |
|--------|---------------|---|------------------------|------------|----------|-----------|-------------------------|
| 1a | Millimeter wave | 60GHz Unlicensed | ≤5 ms | ≤800 Mbit/s | PtP (LOS) | TDD | |
| 1b | | | ≤200 μsec | ≤1Gbps | PtP (LOS) | FDD | |
| 1c | | 70-80GHz Light | ≤200 μsec | ≤2.5 Gbit/s | PtP (LOS) | FDD | |

| | | licensed | | | | |
|---|---|---|---|---|---|---|
| **2a** | Microwave (28-42 GHz) Licensed | ≤200 μsec | ≤1Gbps | PtP (LOS) | FDD | |
| **2b** | | ≤10 ms | ≤1Gbps | PtmP (LOS) | TDD | TDMA |
| **3a** | Sub-6 GHz Unlicensed or licensed | ≤5 ms | ≤500Mbps | PtP (NLoS) | TDD | |
| **3b** | | ≤10 ms | ≤500Mbps (shared) | PtmP (NLoS) | TDD | TDMA |
| **3c** | | ≤5 ms | ≤1 Gbit/s (per client) | PtmP (NLoS) | TDD | SDMA |
| **4a** | Dark Fibre | 5 μs/km × 2 | ≤10 Gbps | PtP | | |
| **4b** | CWDM | 5 μs/km × 2 | ≤10·N Gbps (with N≤8) | Ring | | WDM |
| **4c** | Metro Optical Network | 250 μs | ≤1 Gbps | Mesh/Ring | | Statistical Packet Multiplexing |
| **4d** | PON (Passive Optical Networks) | ≤1 ms | 100M–2.5Gbps | PtmP | | TDM (DL) TDMA (UL) |
| **5** | xDSL | 5-35 ms | 10M–100Mbps | PtP | | |
| **6** | 1 Gigabit Ethernet | ≤200 μs | 1Gbps | PtP | | |

*Table 5: Backhaul Characterisation [16].*

Based on this taxonomy, iJOIN has mapped the existing BH technologies with possible degree of virtualisation on the access network. As described, in Section 4, higher degrees of functional split are associated to stringent latency constraints and high capacity requirements. On the contrary, centralizing only upper layers of the protocol stack does not have an important impact of the bandwidth of the transport network and latency requirements can be notably relaxed. In particular, the BH capacity affects the amount of signalling information that can be exchanged to enable centralised signal processing and/or radio resource management. Additionally, the BH latency affects the CSI/CQI reliability, and interference management schemes may not lead to notable centralisation gains when the transport network induces high latencies. Therefore, the feasibility of a specific degree of virtualisation depends on the latency and the throughput imposed by the BH technology. On the one hand, for wireless solutions, the latency and the achievable throughput depend on the range of frequencies employed, the availability of Line of Sight transmissions, and the topology. On the other hand, for fibre links, the main relevant parameters are the topology and the multiplexing technologies.

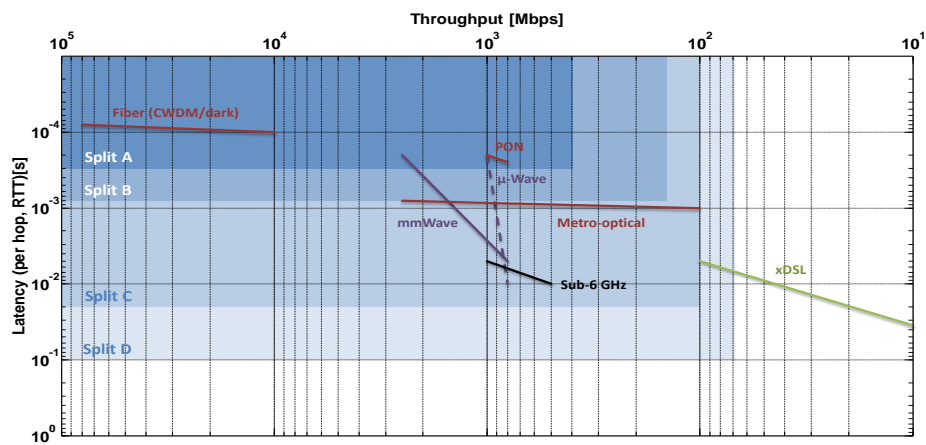

*Figure 17 :Mapping of BH technologies and degree of virtualisation of the RAN [iJOIN D5.3].*

Figure 17 shows the mapping of the BH technologies with different degree of virtualisation, i.e., fully virtualised RAN (split A), PHY upper layer virtualisation (split B), MAC layer (split C), and PDCP and RRC virtualisation (split D).

### 3GPP categories and SPEED-5G focus

The 3GPP initiative categorises backhaul to ideal and non-ideal [20]. Ideal backhaul has very high throughput and very low latency links, such as dedicated point-to-point connections using optical fibre. Non-ideal is the typical backhaul widely used in the market such as xDSL, wireless, and other backhaul technologies like relaying. In this study, we will focus on non-ideal wireless backhaul.

Wireless backhaul technologies typically use micro-wave (MW) and millimetre-wave (mmW) frequencies. Regarding the topology, backhaul systems are designed to support either Point-to-Point (PtP) or Point-to-Multi-Point (PtMP) connectivity. Systems augmented with Software Defined Radio (SDR) capabilities can support both with the same hardware equipment and are programmed to work in either PtP or PtMP mode, depending on the application.

The backhaul network of an operator may combine all topologies and – in some cases – with a single hardware equipment (using SDR). Intelligent wireless backhaul solutions, in particular, providing flexible multi-point (PtP /Relay/PtMP) topologies are of utmost importance in supporting high-capacity, last-mile access and aggregation networks at the macro and small-cell layers.

Figure 18 shows a typical example of a backhaul network with PtP and PtMP configurations for small cells and macro cell. In general, each backhaul equipment is connected to an access base station (BS), be that small cell or macro BS. In the figure, we see a mix of different topologies. PtP systems provide a connection between two nodes and can also be used as relays (repeaters) in a multi-hop backhaul network. PtMP systems use a hub that connects to a number of terminals. The hub constitutes an aggregation point for the backhaul network when combined with more hubs can cover wider angles. For example, four hubs covering a 90$^{\circ}$ sector each can cover a 360$^{\circ}$ area.

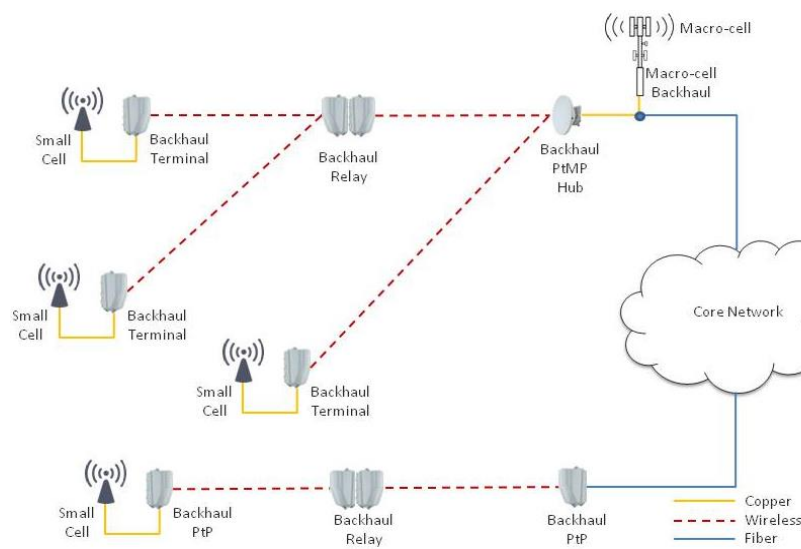

*Figure 18: Backhaul Network with PtP and PtMP Configurations for Small Cells and Macro cell*

LTE/4G Radio Access Networks (RANs) currently evolve to Heterogeneous Networks (HetNets) by seamlessly integrating macro and small cells of various technologies. Selecting and utilizing the perfect mix of backhaul solutions is a major challenge when deploying a new RAN. Packet wireless technologies are essential elements of HetNets, offering excessive IP capacity and extending the advantages of PtP technology to street-level backhaul by using PtMP technology.

PtP and PtMP systems may require Line-of-Sight (LOS) depending on their frequency of operation. LOS is required for higher frequencies (mmW), while lower frequencies can work in Non-Line-of-Sight (NLOS) or LOS. In addition, depending on local regulations, the higher the frequency selected the larger the channel bandwidth available and hence the data rate. On the contrary, lower frequency signals travel further hence the distance between the two ends can be longer.

A PtMP topology is selected in cases where one hub connects multiple terminals in a cost effective manner. Various software planning tools are used to determine feasibility of potential connections using topographic data as well as link budget simulation. In general, performance of PtMP links is lagging behind this of the PtP ones due to the lower frequencies used, the multi-access protocol overhead and the link sharing among terminals.

Our study in this section focuses on PtMP systems and the improvement of their performance towards 5G KPIs, since their cost advantage over fibre and PtP wireless systems as well as their installation ease, will continue to render them important to the operators.

Finally, it must be noted that in 5G, both *D2D communication* and *smart caching* call for an architectural redefinition where the centre of gravity moves from the network core to the periphery (devices, local wireless proxies, relays). Based on these trends, the UE-centric network vision calls for an evolution of the "old" cell-centric architecture into a device-centric one: a given device (human or machine) should be able to communicate by exchanging multiple information flows through several possible sets of heterogeneous nodes. In other words, the set of network nodes providing connectivity to a given device and the functions of these nodes in a particular communication session should be tailored to that specific device, service and session. With the above in mind, the actual usage of the backhaul network depends both on the type service and the data locality at a given point in time. The study provided in this deliverable for backhaul performance metrics applies, of course, to the services that actually use it.

### PtMP backhaul link latency

Figure 19 shows PtMP links between a hub, H, and some terminal nodes, $T_1$, $T_2$ to $T_N$. PtMP wireless communication is established between each terminal and the hub. In the downstream direction, information is broadcasted to all terminals with each keeping only the information addressed to it. In the upstream direction Time Division Multiple Access (TDMA) is used from the terminals to transmit over the same frequency.

Inside each node, roughly two functions are implemented:

- Functions regarding layer 1 (L1): hardware implemented functions regarding the transmission and reception over their intervening medium, i.e. the air. L1 is proprietary. It differs from the PtP mode in that multiple access communication must be supported.
- Functions regarding layer 2 (L2): processor (µP) assisted functions regarding Ethernet switching and bridging.

The latency in the DS direction per pair, $T_{n\{DS\}}$, n=1,2,..,N, is the sum of the following:

- The time to process the information inside node H, i.e. to pass through L2 and L1 functions, $T_{pH\{DS\}}$

- The transmission time over the air, $T_{xmn\{DS\}}$. This differs per pair and depends on the distance between the hub and the terminal.

- The time to process the information inside node $T_n$, i.e to pass through L1 and L2 functions, $T_{pTn\{DS\}}$. It depends on QoS parameters, e.g. the traffic each terminal is required to process and the priority of each frame.

The latency in the US direction per pair, $T_{n\{US\}}$, n=1,2,..,N, is the sum of the following:

- The time to process the information inside node $T_n$, i.e to pass through L2 and L1 functions, $T_{pTn\{US\}}$. It again depends not only on QoS parameters, but also the time slot assigned to the node for transmission.

- The transmission time over the air, $T_{xmn\{US\}}$. This differs per pair and depends on the distance between the hub and the terminal.

- The time to process the information inside node H, i.e. to pass through L2 and L1 functions, $T_{pH\{US\}}$.

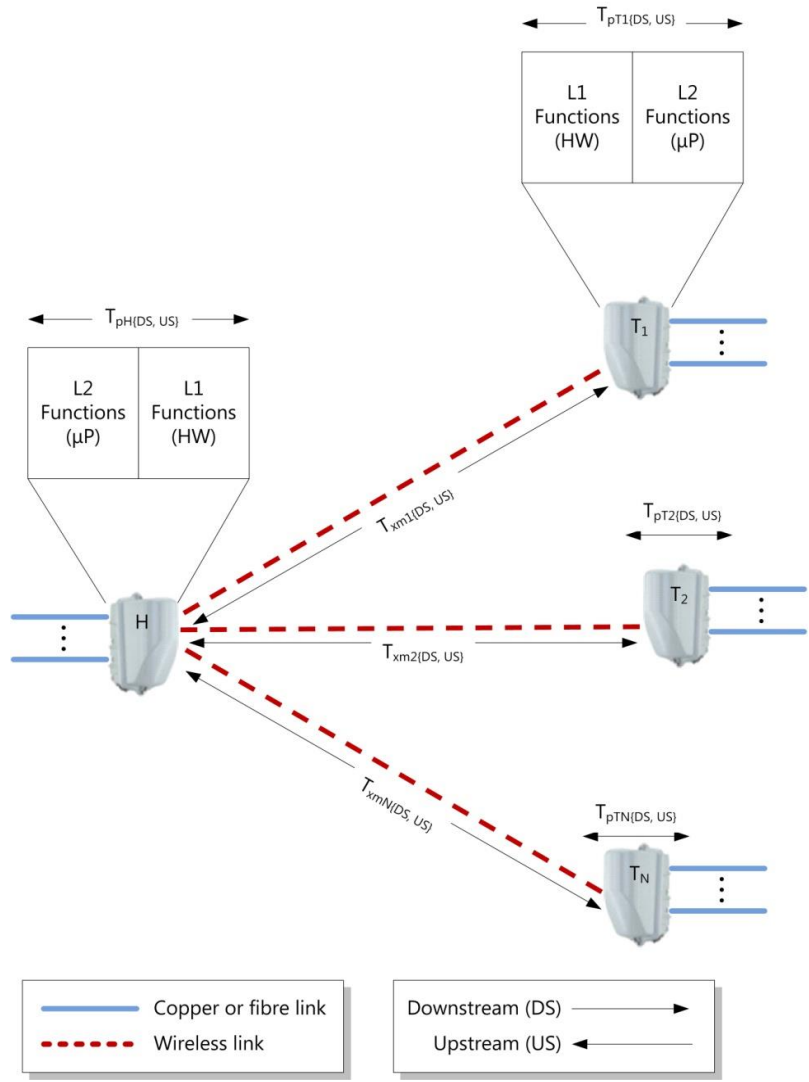

*Figure 19: Latency in PtMP links*

To conclude, latency in PtMP mode depends both on the direction, DS or US, and on the pair, {H,T$_n$}:

$$T_{n\{DS\}} = T_{pH\{DS\}} + T_{xmn\{DS\}} + T_{pTn\{DS\}}$$

$$T_{n\{US\}} = T_{pTn\{US\}} + T_{xmn\{US\}} + T_{pH\{US\}}$$

The target value for one way latency compared to 5G Use Cases under consideration and 3GPP release 12 is shown in Table 6.

| Target latency in PtMP backhaul (one way) | |
|---|---|
| $\simeq$ 1ms | |
| **5G use case covered** | |
| Massive IoT communication | Orders of seconds or more (E2E) |
| High-Speed mobility | 10ms (E2E) |
| Broadband access | 10ms (E2E) |
| Ultra-reliable communications (*) | 10ms (E2E) |
| **Wireless backhaul latency in 3GPP rel 12 (one way)** | |
| 5-35ms | |

*Table 6: PtMP backhaul latency target wrt 5G use cases and 3GPP rel 12 [25]*

(*) For the ultra-reliable communications use case family, the following sub-case is covered, as listed by NGMN:

- Ultra-high availability & reliability (DL: 10 Mbps, UL: 10 Mbps, E2E latency: 10 ms)

All the other Use Cases are covered as listed in deliverable D3.1 "Value chain analysis and system design".

## PtMP backhaul link data rate

Figure 20 shows the same PtMP example as presented in the latency study, using notation for the data rate metric. The data rate in the DS direction per pair, $R_{n\{DS\}}$, n=1,2,..,N, is the minimum of the following:

- The processor throughput of the hub, $R_{\mu PH\{DS\}}$. This basically depends on the selected processor. Most modern network processors include state-of-the-art technologies, such as hardware accelerators for functions like traffic classification, policing, shaping etc, in order to Speed up the frame processing to the maximum possible.

- The data rate of the physical medium, i.e. the air, per pair, $R_{Ln\{DS\}}$, n=1,2,..,N. This is actually achieved by the L1 functions during the initialisation of the system. The actual data rate that can be achieved by the link is a function of both the L1 implementation and the physical medium, and depends on a number of parameters such as: the air interface selected, the channel bandwidth, the number of antennas (e.g. MIMO), the distance between the nodes, the weather conditions, physical obstacles that may intervene, earth curvature and more. Moreover, overall traffic QoS settings (scheduling) may limit the actual data rate achieved per pair.

- The processor throughput of each terminal, $R_{\mu PTn\{DS\}}$. The same comments apply here as in the hub processor.

In the US direction, the data rate per pair, $R_{n\{US\}}$, n=1,2,..,N, is the minimum of the following, with the same comments applying here as in the DS case.:

- The processor throughput of each terminal, $R_{\mu PTn\{US\}}$.

- The data rate of the physical medium, i.e. the air, per pair, $R_{Ln\{US\}}$, n=1,2,..,N.

- The processor throughput of the hub, $R_{\mu PH\{DS\}}$.

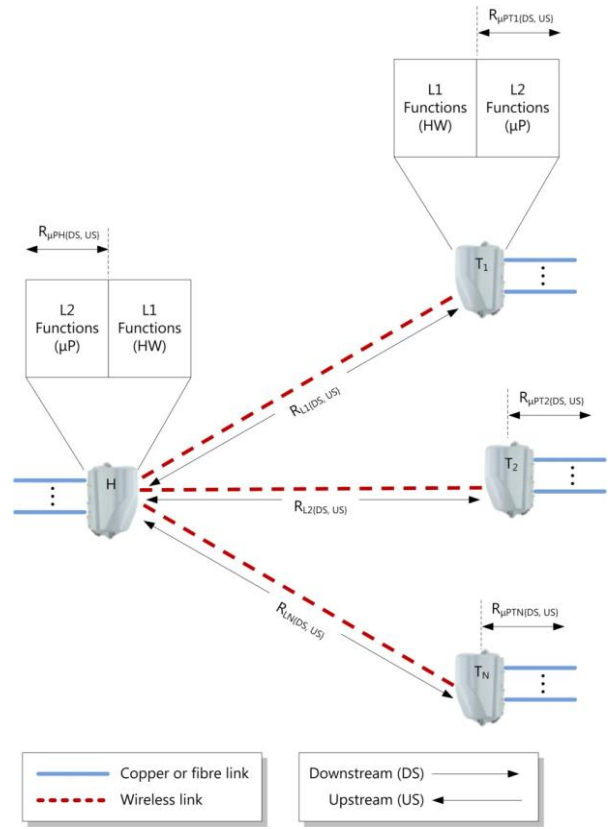

*Figure 20: Data rate in PtMP backhaul links*

To conclude data rate in PtMP mode depends both on the direction, DS or US, and on the pair, $\{H, T_n\}$:

$$R_{n\{DS\}} = \min\{ R_{\mu H\{DS\}}, R_{Ln\{DS\}}, R_{\mu PTn\{DS\}} \}$$

$$R_{n\{US\}} = \min\{ R_{\mu H\{US\}}, R_{Ln\{US\}}, R_{\mu PTn\{US\}} \}$$

The target values data rate latency compared to 5G Use Cases under consideration and 3GPP release 12 is shown in Table 7.

| Target data rate in PtMP backhaul link | |
|---|---|
| $\simeq$ 2.5 Gbps (DS) | |
| $\simeq$ 2.0 Gbps (US) | |
| **5G Scenario covered** | |
| Massive IoT communication | From tens to hundreds of Kbps |
| High-Speed mobility | 50Mbps (DS) 25Mbps (US) |
| Broadband access (*) | 50 Mbps (DS) 25Mbps (US) |
| Ultra-reliable communications | Several Kbps |
| **Wireless backhaul data rate in 3GPP rel 12** | |
| 10Mbps – 100Mbps typical, maybe up to Gbps range | |

*Table 7: PtMP backhaul latency target wrt 5G use cases and 3GPP rel. 12 [25]*

(*) For the broadband use case family, the following sub-cases are well covered, as listed by NGMN:

- Broadband access in a crowd (DL: 25 Mbps, UL: 50 Mbps, 10 ms)

- 50+ Mbps everywhere (DL: 50 Mbps, UL: 15 Mbps, 10 ms)

- Ultra-low cost broadband access for low ARPU areas (DL: 10 Mbps, UL: 10 Mbps, 10 ms)

All the other Use Cases are covered as listed in deliverable D3.1 "Value chain analysis and system design".

## Network synchronisation over PtMP backhaul

Network synchronisation is and will continue to be important to telecom operators for them to provide high quality services to their customers. The mobile network evolution to LTE and future planning for 5G networks and services has generated an increased and pressing need for the delivery of accurate synchronisation. Apart from the need for these networks to provide ever-increasing data rates and lower network latencies, accurate **end-to-end** synchronisation schemes are needed to support new features. Examples include the coordination between base stations when delivering broadcast video and the avoidance of interference between macro and small cell base stations. Inaccurate - or no -synchronisation can result in the following issues:

- Poor call quality and noticeable "clicks" during phone calls

- Problems in call setup, termination and management

- Problems in cell coordination especially during call handover

- Reduced data rates

- Screen freezes and sound issues in video transmissions

- Partial or total data traffic disruption

Different radio technologies and features have different synchronisation requirements. These can be categorised into two main types: frequency synchronisation and phase synchronisation. Many different options exist to provide frequency synchronisation although fewer exist that can reliably deliver the required accuracy and stability for phase synchronisation. Our work will focus on Synchronous Ethernet (SyncE) for frequency synchronisation and the Precision Time Protocol (PTP) IEEE 1588v2 for phase synchronisation. Both options will be adopted by the PtMP system under investigation to provide a complete synchronisation solution.

The main principle of SyncE is to synchronise the interfaces of a network system to a clock derived from an Ethernet port. The principle and its difference with plain Ethernet is shown in Figure 21.

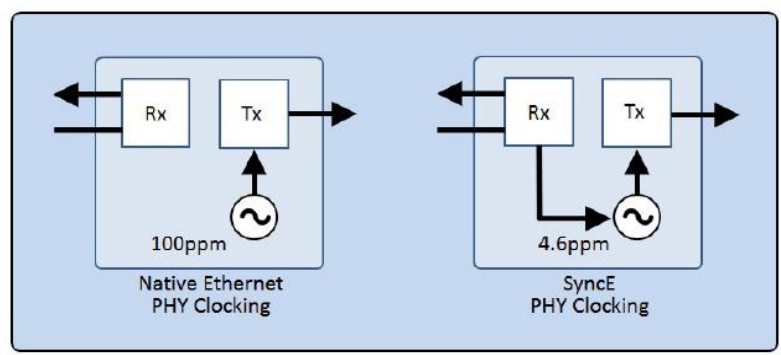

*Figure 21: Native versus Synchronous Ethernet*

SyncE is standardised by ITU-T, in cooperation with IEEE. Within this framework the following recommendations have been published:

- ITU-T G.8261. It defines architecture and accuracy related subjects in SyncE networks.

- ITU-T G.8262. It defines the attributes of the SyncE equipment clocks. These are known as Ethernet Equipment Clocks (EEC). While IEEE 802.3 requires that the clock accuracy is ±100ppm, the EEC accuracy is required to be ±4.6ppm.

- ITU-T G.8264. It defines the Ethernet Synchronisation Messaging Channel (ESMC), extending the ITU-T G.707 standard. The message used in this channel, Synchronisation Status Message (SSM), contains a quality indication of the clock that a synchronisation chain is locked to.

Precision Time Protocol (PTP) is standardised as IEEE1588 (latest version 2) and officially named "Standard for a Precision Clock Synchronisation Protocol for Networked Measurement and Control Systems". The main idea is that a master device has a precise clock and all other compatible devices in the chain can synchronise themselves to it through packet exchange. **Figure 22** shows an example with one master, three slaves and a boundary device.

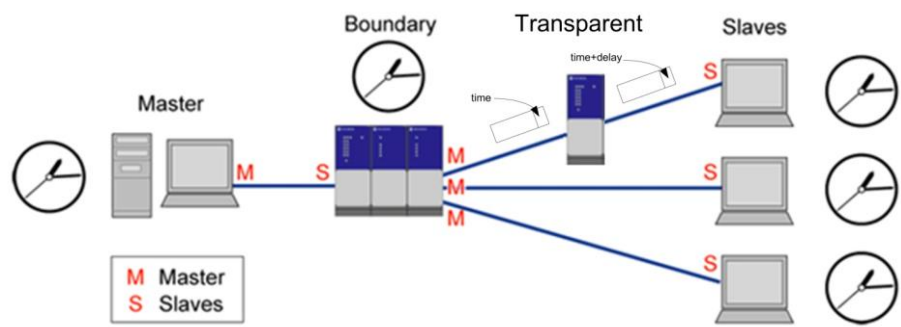

*Figure 22: Example of devices in a 1588v2 chain*

According to the standard a device can be one of the following:

- <u>Ordinary Clock</u>: a device in the synchronisation chain which has one port that can be either a 1588v2 master or slave. These are devices at the edges of the chain

- <u>Boundary Clock</u>: a device in the synchronisation chain that has at least two ports that can be either a 1588v2 master or a slave. Such a device has a slave clock that recovers the IEEE1588v2 from the slave port. It is then used to drive the clock of its master clock that feeds the next nodes.

- <u>Transparent Clock</u>: a device in the chain with at least two ports that are neither IEEE1588v2 master nor slave. It operates as a transparent bridge between the two ports, modifying and forwarding the IEEE1855v2 messages by adding the delay the latter faced in the device (because of buffering and transmission times). The modification is done by adding this delay to the Correction Field of the message. This way a master or slave is able to consider this delay in its computations.

- <u>Management Node</u>: a device that configures and monitors the rest synchronisation devices.

Our study focuses on providing high-accuracy synchronisation over the PtMP wireless backhaul segment, which supports both synchronisation methods. Figure 23 depicts an example of a mobile network supporting these methods. On the left edge, a Synchronisation Supply Unit (SSU) uses a high-accuracy Primary Reference Source (PRS) as a "good" reference clock to feed the chain with. The SSU is considered both a SyncE and 1588v2 master. In the backhaul network, the left port of the Hub is considered a SyncE slave which recovers the SyncE clock from the previous stage. The clock is transmitted over the air to the Terminals ($T_1$, $T_2$, .., $T_N$) which use it to drive their ports on the right which are thus considered SyncE masters. In the example the final SyncE slaves are the cells attached to the Terminals. In the IEEE 1588v2 chain the SSU is the master and the cells on the other edge are the slaves. Each PtMP pair, i.e. {H, T1}, {H, T2}, .. , {H, TN}, is a – distributed – Transparent Clock (TC). The PTP message is modified by each TC by adding to the Correction Field the time the message spent between its entrance to the Hub and its exit from the Terminal.

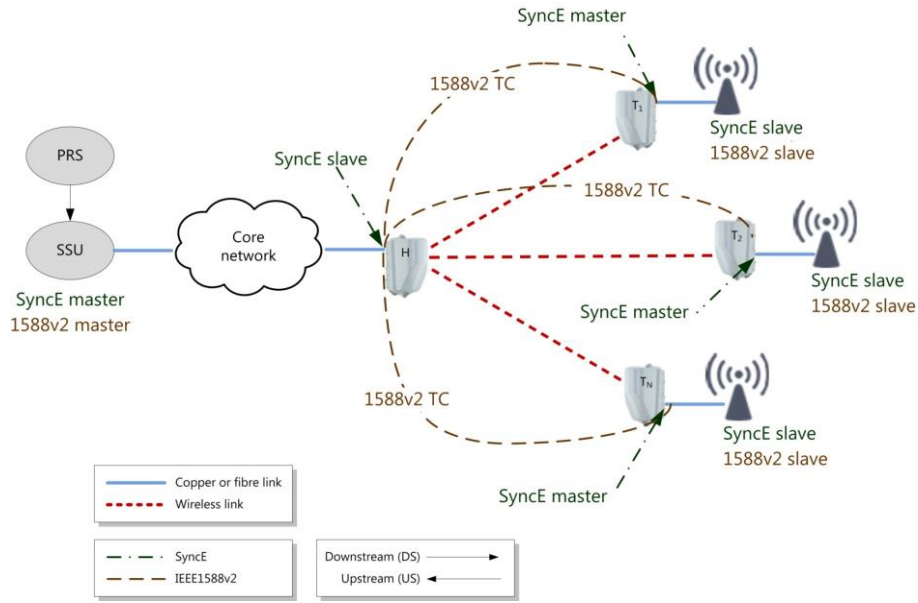

*Figure 23: SyncE and IEEE1588v2 TC in PtMP backhaul*

| Sync method | Clock mode | Metric | Value | | Standard |
|---|---|---|---|---|---|
| SyncE | Locked or holdover | MTIE | **MTIE limit [ns]** / **Observation interval τ [s]**: 40 / 0.1 < τ ≤ 1; 40 τ$^{0.1}$ / 1 < τ ≤ 100; 25.25 τ$^{0.2}$ / 100 < τ ≤ 1000 | | ITU-T G.8262/Y.1362 |
| | | TDEV | **TDEV limit [ns]** / **Observation interval τ [s]**: 3.2 / 0.1 < τ ≤ 25; 0.64 τ$^{0.5}$ / 25 < τ ≤ 100; 6.4 / 100 < τ ≤ 1000 | | |
| | Holdover or free-run | Output frequency accuracy | within ±4.6ppm | | |
| IEEE 1588v2 | - | CF accuracy | ≤ 80 ns | | Not yet defined. Under work in ITU-T G8273.3 standard |

*Table 8: Network synchronisation metrics for PtMP backhaul system wrt to ITU standards [26][27]*

Table 8 summarises the synchronisation metrics and their target values. In SyncE, the noise generation of an EEC represents the amount of phase noise produced at the output when there is an ideal input reference signal or the clock is in holdover state. A suitable reference, for practical testing purposes, implies a performance level of at least ten times more stable than the output requirements. The ability of the clock to limit this noise is described by its frequency stability. The maximum time interval error (MTIE) and time deviation (TDEV) are useful means for characterisation of noise generation performance. MTIE and TDEV are measured through an equivalent 10-Hz, first-order, low-pass measurement filter, at a maximum sampling time $\tau_0$ of 1/30 seconds. The minimum measurement period for TDEV is twelve times the integration period ($T = 12\tau$). The nested tables of the target values of MTIE and TDEV are shown as graphs in Figure 24 and Figure 25 respectively.

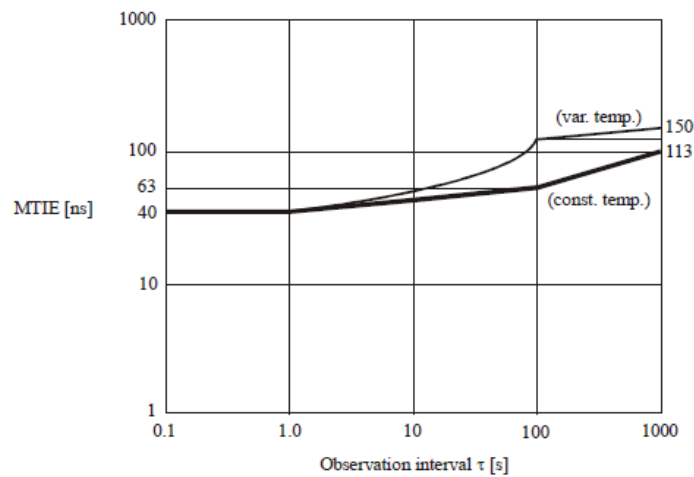

*Figure 24: MTIE wrt observation interval [26]*

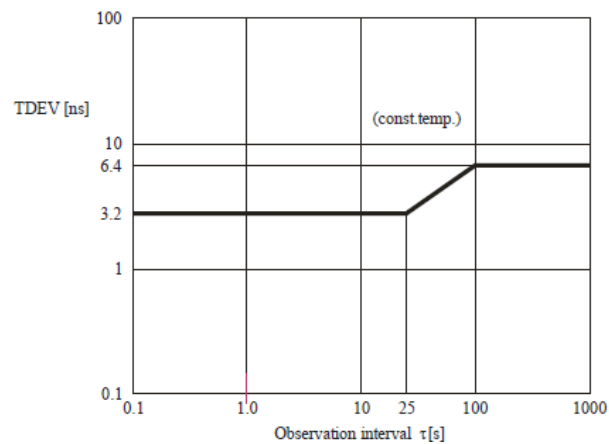

*Figure 25: TDEV wrt observation interval [26]*

## Resource balancing and high availability through redundancy in PtMP backhaul

Resource balancing and high availability in a PtMP cluster will be addressed through redundancy. Figure 26 presents the idea. The hub of the cluster consists of two equivalent nodes (H1 and H2), operating in a 1:1 mode. Each terminal node is linked to only one of the hub nodes decided by an intelligent application which monitors the hub nodes. When one of the hub nodes fails, the application will restructure the cluster so that all terminal nodes are assigned to the operational one, as Figure 27 shows.

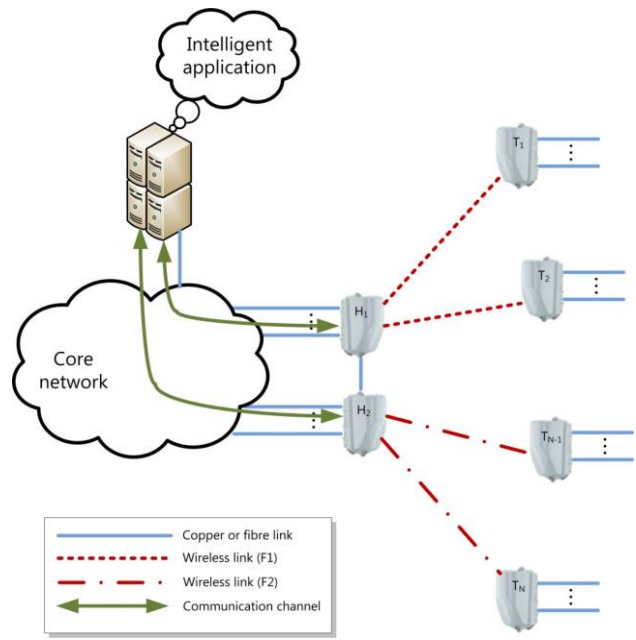

*Figure 26: Redundancy in PtMP cluster*

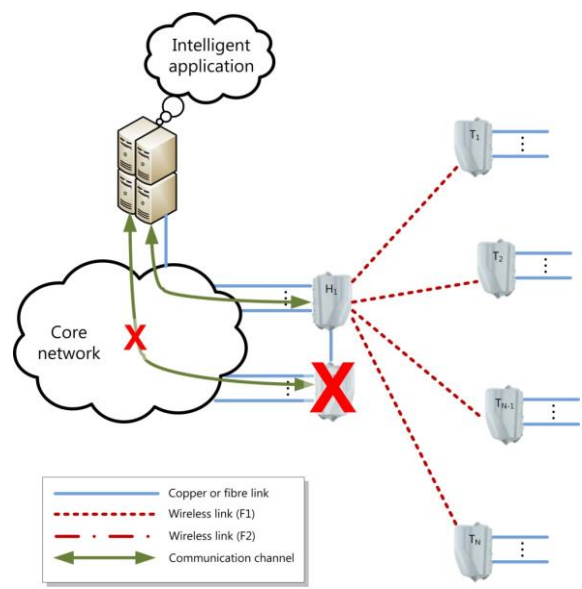

*Figure 27: Failure in a PtMP hub node*

By using a pair of hub nodes, the availability of the cluster is increased. In what follows, we use Reliability Theory to calculate the improvement.

We consider the following parameters for each node:

**Mean Time Between Failures (MTBF)**. The average time between failures of a node. It is the average time the manufacturer estimates before a failure occurs in the node.

**Mean Time To Repair (MTTR)**. The time taken to repair a failed node. In an operational node, repair generally means replacing the hardware module. For a telecom operator, MTTR depends on where the node is located with respect to the technicians' base and can greatly vary from several minutes (simple on-site replacement) to even days if the node is placed in a location where travelling is required to reach it.

**Availability (A)**. It is the percentage of time that the node is operational. It is defined as

$$A = \frac{\text{MTBF}}{\text{MTBF} + \text{MTTR}}$$

**Downtime per year (D)**. It is the amount of time in a year that the node is not operating. It is defined as

$$D = (1 - A) \times 1 \text{ year}$$

Availability is usually referenced as "nines", indicating the number of continuous "nines" in the percentage format. The following table shows a few examples of how availability and downtime are related.

| Availability | Downtime (per year) |
|---|---|
| 90% (1-nine) | 36.5 days |
| 99% (2-nines) | 3.65 days |
| 99.9% (3-nines) | 8.76 hours |
| 99.99% (4-nines) | 52 minutes |
| 99.999% (5-nines) | 5 minutes |
| 99.9999% (6-nines) | 31 seconds |

*Table 9: Availability and downtime examples*

In our study, we also take into account that all nodes have common hardware. Their role (hub or terminal) is defined by their software. Hence, each node has the same availability with each other, A.

### Availability of a sector with one hub node

Let us consider a scenario where a PtMP cluster with N terminals covers a sector and we mark it as available if at least the hub and M of the terminals are available (M ≤ N). The reliability block diagram of this system is presented in Figure 28. The hub and the terminals sub-cluster are placed in series because in order for the whole cluster to be available both must be available.

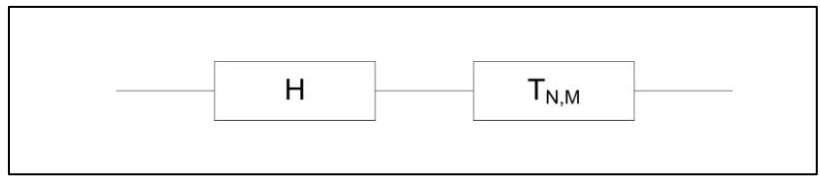

*Figure 28: Reliability block diagram of a PtMP cluster with one hub node*

The cluster availability is defined as

$$A_C = A_H \times A_{T(N,M)}$$

$A_H$ is the availability of the hub node, i.e. A.

$A_{T(N,M)}$ is the availability of a sub-cluster of N terminals, that is considered to be available when at least M of them is available. It is calculated by the following formula, where A is the availability of each node:

$$A_{T(N,M)} = \sum_{i=0}^{N-M} \binom{N}{i} \times A^{(N-i)} \times (1-A)^i$$

In each sum term of the above formula, $i$ is the number of non-available nodes and $N\text{-}i$ is the number of the available ones. So, $A^{(N-i)}$ is the availability of the $N\text{-}i$ terminals, $(1-A)^i$ is the failure probability of the rest. Hence the formula calculates the probability of $N\text{-}i$ terminals being operational and $i$ not being operational for all the combinations of $i$ terminals from $N$, i.e $\binom{N}{i}$.

### Availability of sector with two hub nodes

In this subsection we consider the same scenario as before, with two hub nodes (hub sub-cluster). When the hub nodes are both healthy they operate in 1:1 mode. When one of them fails, all its terminals switch to the healthy one by the intelligent application. The hub sub-cluster is considered available when at least one of its nodes is available. The reliability block diagram of this system is presented in Figure 29. The two hub nodes are placed in parallel because one of them is required to be available. Then the sub-cluster of the terminals is placed in series because eventually we need to also have M terminals available.

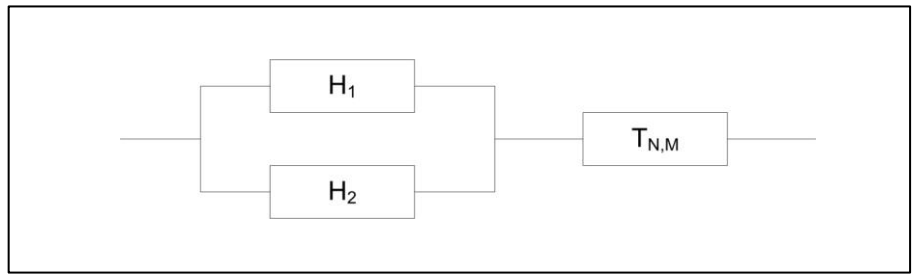

*Figure 29: Reliability block diagram of a PtMP cluster with two hub nodes*

Thus, the cluster availability is

$$A_C = A_H \times A_{T(N,M)}$$

$A_H$ is the combined availability of the parallel pair of hub nodes, i.e. 1 - (both parts are unavailable). So,

$$A_H = 1-(1-A)^2$$

The implication of the above equation is that the combined availability of two components in parallel is always much higher than the availability of its individual components. Considering the system in Figure 29, the table below shows an example of how availability and downtime are increased when considering H1 and H2.

| Component | Availability | Downtime (per year) |
|---|---|---|
| H | 99% (2-nines) | 3.65 days |
| H1 and H2 in parallel | 99.99% (4-nines) | 52 minutes |

*Table 10: How redundancy increases availability*

$A_{T(N,M)}$ is the availability of a sub-cluster of N terminals, that is considered to be available when at least M of them is available. It is calculated by the formula:

$$A_{T(N,M)} = \sum_{i=0}^{N-M} \binom{N}{i} \times A^{(N-i)} \times (1-A)^i$$

 

## Availability calculation and comparison table

It is not possible to give a global number of availability as it is dependent on the MTBF, MTTR and the scenario under discussion. In the following table we provide calculations for MTBF = 50 years (438000 hours) and MTTR = 3 hr, 12 hr, 24 hr and 72 hr, based on the mathematical formulas of the above scenarios. Also a sector of N = 3 and M = 1 is modeled, i.e. we choose to consider it available when at least one of the 3 terminals and a hub node are available.

| MTTR | | With one hub node | | With two hub nodes | |
|---|---|---|---|---|---|
| | | Availability | Downtime (per year) | Availability | Downtime (per year) |
| 3 hrs | | 99,9993150731844% 5-nines | 3,6 min | 99,9999999953087% 10-nines | 1,5 msec |
| 12 hrs | | 99,9972603490295% 4-nines | 14,4 min | 99,9999999249411% 9-nines | 23,7 msec |
| 24 hrs | | 99,9945208481563% 4-nines | 28,8 min | 99,9999996997725% 9-nines | 94,7 msec |
| 72 hrs | | 99,9835643451431% 3-nines | 86,4 min | 99,9999972982487% 7-nines | 852 msec |

*Table 11: PtMP sector availability and downtime calculations*

It is clearly shown that redundancy of the hub node greatly increases the availability of the system. In the best case (MTTR=3hr), availability is increased from "5 nines" to "10 nines". In the worst case (MTTR=3days), availability is increased from just "3 nines" to "7 nines".